\documentclass[preprint,showpacs,preprintnumbers,amsmath,amssymb]{revtex4}
\usepackage{epsfig}
\usepackage{graphicx}% Include figure files
\usepackage{dcolumn}% Align table columns on decimal point
\usepackage{bm}% bold math

\def\beq{\begin{equation}}
\def\eeq{\end{equation}}
\def\eeqn{\end{equation}}
\newcommand\iden{\leavevmode\hbox{\small1\normalsize\kern-.33em1}}

\newcommand{\bea} {\begin{eqnarray}}
\newcommand{\eea} {\end{eqnarray}}

\newcommand{\nn}{\nonumber}
\let\jnfont=\rm
\def\NPB#1,{{\jnfont Nucl.\ Phys.\ B }{\bf #1},}
\def\PLB#1,{{\jnfont Phys.\ Lett.\ B }{\bf #1},}
\def\EPJC#1,{{\jnfont Eur.\ Phys.\ Jour.\ C }{\bf #1},}
\def\PRD#1,{{\jnfont Phys.\ Rev.\ D }{\bf #1},}
\def\PRL#1,{{\jnfont Phys.\ Rev.\ Lett.\ }{\bf #1},}
\def\MPLA#1,{{\jnfont Mod.\ Phys.\ Lett.\ A }{\bf #1},}
\def\JPG#1,{{\jnfont J.\ Phys.\ G }{\bf #1},}
\def\CTP#1,{{\jnfont Commun.\ Theor.\ Phys.\ }{\bf #1},}
\def\JHEP#1,{{\jnfont JHEP \ }{\bf #1},}
\def\NPPS#1,{{\jnfont Nucl.\ Phys.\ Proc.\ Suppl.\ }{\bf #1},}
\def\CPC#1,{{\jnfont Comput.\ Phys.\ Commun.\ }{\bf #1},}
\def\CPL#1,{{\jnfont Chin.\ Phys.\ Lett. }{\bf #1},}
\def\APPB#1,{{\jnfont Acta\ Phys.\ Polon.\ B }{\bf #1},}

\def\lsim{\raise0.3ex\hbox{$<$\kern-0.75em\raise-1.1ex\hbox{$\sim$}}}
\def\gsim{\raise0.3ex\hbox{$>$\kern-0.75em\raise-1.1ex\hbox{$\sim$}}}

\begin{document}

\title{\ \\[10mm] Status of the aligned two-Higgs-doublet model confronted with the Higgs data}

\author{Lei Wang, Xiao-Fang Han}

\affiliation{
Department of Physics, Yantai University, Yantai 264005, PR China
\vspace{0.5cm} }

%---------------------------------------------------------------------------

\begin{abstract}
Imposing the theoretical constraints from vacuum stability,
unitarity and perturbativity as well as the experimental constraints
from the electroweak precision data, flavor observables and the
non-observation of additional Higgs at collider, we study the
implications of available Higgs signals on a two-Higgs-doublet model
with the alignment of the down-type quarks and charged lepton Yukawa
coupling matrices. Compared to the four traditional types of
two-Higgs-doublet models, the model has two additional mixing angles
$\theta_d$ and $\theta_l$ in the down-type quark and charged lepton
Yukawa interactions. We find that the mixing angle $\theta_d$ can
loose the constraints on $\sin(\beta-\alpha)$, $\tan\beta$ and
$m_{H^{\pm}}$ sizably. The model can provide the marginally better
fit to available Higgs signals data than SM, which requires the
Higgs couplings with gauge bosons, $u\bar{u}$ and $d\bar{d}$ to be
properly suppressed, and favors (1 $<\theta_d<$ 2, 0.5 $<\theta_l<$
2.2) for $m_h=$ 125.5 GeV and (0.5 $<\theta_d<$ 2, 0.5 $<\theta_l<$
2.2) for $m_H=$ 125.5 GeV. However,  these Higgs couplings are
allowed to have sizable deviations from SM for ($m_h=$ 125.5 GeV,
125.5 $\leq m_H\leq$ 128 GeV) and (125 GeV $\leq m_h\leq$ 125.5 GeV,
$m_H=$ 125.5 GeV).
\end{abstract}

\maketitle

\section{Introduction}
The CMS and ATLAS collaborations have announced the observation of a
scalar around 125 GeV \cite{cmsh,atlh}, which is supported by the
Tevatron search \cite{tevh}. The properties of this particle with
large experimental uncertainties are well consistent with the SM
Higgs boson, which will give the strong constraints on the effects
of new physics.

One of the simplest extension of the SM is obtained by adding a
second $SU(2)_L$ Higgs doublet \cite{2hdm}. The two-Higgs-doublet
model (2HDM) has very rich Higgs phenomenology, including two
neutral CP-even Higgs bosons $h$ and $H$, one neutral pseudoscalar
$A$, and two charged Higgs $H^{\pm}$. Further, the couplings of the
CP-even Higgs bosons can deviate from SM Higgs boson sizably.
Therefore, the observed signal strengths of the Higgs boson and the
non-observation of additional Higgs can give the strong implications
on the 2HDMs. The 2HDMs generically have tree-level flavor changing
neutral currents (FCNC), which can be forbidden by a discrete
symmetry. There are four types for 2HDMs, which are typically called
the Type-I \cite{i-1,i-2}, Type-II \cite{i-1,ii-2}, Lepton-specific,
and Flipped models \cite{xy-1,xy-2,xy-3,xy-4,xy-5,xy-6} according to
their different Yukawa couplings.  In light of the recent Higgs
data, there have been various studies on these 2HDMs over the last
few months
\cite{2h-1,2h-2,2h-3,2h-4,2h-5,2h-6,2h-7,2h-afb,2h-8,2h-9,2h-11,2h-10,2h-12,2h-13}.

In this paper, we focus on a two-Higgs-doublet model that allows
both doublets to couple to the down-type quarks and charged leptons
with aligned Yukawa matrices ( A2HDM) \cite{2h-11,a2hm-1}. Also
there is no tree-level FCNC in this model. Compared to the above
four types of 2HDMs, there are two additional mixing angles in the
Yukawa couplings of the down-type quarks and charged leptons. This
model can be mapped to the four types of 2HDMs for the two angles
are taken as specific values. There are also some works on the Higgs
properties in the A2HDM after the discovery of Higgs boson
\cite{2h-11,2h-10,a2hw-1,a2hw-2,a2hw-3,a2hw-4,a2hw-6,a2hw-5}. After
imposing the theoretical constraints from vacuum stability,
unitarity and perturbativity as well as the experimental constraints
from the electroweak precision data, flavor observables and the
non-observation of additional Higgs at collider, we study the
implication of the latest Higgs signals data on the A2HDM.

Our work is organized as follows. In Sec. II we recapitulate the
A2HDM. In Sec. III we introduce the numerical calculations. In Sec.
IV, we discuss the implications of the available Higgs signals on
the A2HDM after imposing the theoretical and experimental
constraints. Finally, we give our conclusion in Sec. V.

\section{aligned two-Higgs-doublet model}
The general Higgs potential is written as \cite{2h-poten}
\begin{eqnarray} \label{V2HDM} \mathrm{V} &=& m_{11}^2
(\Phi_1^{\dagger} \Phi_1) + m_{22}^2 (\Phi_2^{\dagger}
\Phi_2) - \left[m_{12}^2 (\Phi_1^{\dagger} \Phi_2 + \rm h.c.)\right]\nonumber \\
&&+ \frac{\lambda_1}{2}  (\Phi_1^{\dagger} \Phi_1)^2 +
\frac{\lambda_2}{2} (\Phi_2^{\dagger} \Phi_2)^2 + \lambda_3
(\Phi_1^{\dagger} \Phi_1)(\Phi_2^{\dagger} \Phi_2) + \lambda_4
(\Phi_1^{\dagger}
\Phi_2)(\Phi_2^{\dagger} \Phi_1) \nonumber \\
&&+ \left[\frac{\lambda_5}{2} (\Phi_1^{\dagger} \Phi_2)^2 + \rm
h.c.\right] + \left[\lambda_6 (\Phi_1^{\dagger} \Phi_1)
(\Phi_1^{\dagger} \Phi_2) + \rm h.c.\right] \nonumber \\
&& + \left[\lambda_7 (\Phi_2^{\dagger} \Phi_2) (\Phi_1^{\dagger}
\Phi_2) + \rm h.c.\right].
\end{eqnarray}
We focus on the CP-conserving model in which all $\lambda_i$ and
$m_{12}^2$ are real. Further, we assume $\lambda_6=\lambda_7=0$,
which also facilitates the comparison to the four traditional types
of 2HDMs. The two complex scalar doublets have the hypercharge $Y =
1$,
\begin{equation}
\Phi_1=\left(\begin{array}{c} \phi_1^+ \\
\frac{1}{\sqrt{2}}\,(v_1+\phi_1^0+ia_1)
\end{array}\right)\,, \ \ \
\Phi_2=\left(\begin{array}{c} \phi_2^+ \\
\frac{1}{\sqrt{2}}\,(v_2+\phi_2^0+ia_2)
\end{array}\right).
\end{equation}
Where $v_1$ and $v_2$ are the electroweak vacuum expectation values
(VEVs) with $v^2 = v^2_1 + v^2_2 = (246~\rm GeV)^2$. The ratio of
the two VEVs is defined as usual to be $\tan\beta=v_2 /v_1$. After
spontaneous electroweak symmetry breaking, the physical scalars are
two neutral CP-even $h$ and $H$, one neutral pseudoscalar $A$, and
two charged scalar $H^{\pm}$. These scalars are also predicted in
the Higgs triplet models \cite{htm1,htm2,htm3}.

The Yukawa interactions of the Higgs doublets with the SM fermions
can be given by \bea - {\cal L} &=& y_u\,\overline{Q}_L \, \tilde{{
\Phi}}_2 \,u_R +\,y_d\, \overline{Q}_L\,(\cos{\theta_d}\,{\Phi}_1
\,+\, \sin{\theta_d}\,
{\Phi}_2) \, d_R   \nonumber \\
&&\hspace{-3mm}+ \, y_l\,\overline{l}_L \, (\cos{\theta_l}\,{\Phi}_1
\,+\, \sin{\theta_l}\, {\Phi}_2)\,e_R \,+\, \mbox{h.c.}\,, \eea
where $Q^T=(u_L\,,d_L)$, $L^T=(\nu_L\,,l_L)$, and
$\widetilde\Phi_2=i\tau_2 \Phi_2^*$. $y_u$, $y_d$ and $y_\ell$ are
$3 \times 3$ matrices in family space. $\theta_d$ and $\theta_l$
parameterize the two Higgs doublets couplings to down-type quarks
and charged leptons, respectively. Where a freedom is used to
redefine the two linear combinations of $\Phi_1$ and $\Phi_2$ to
eliminate the coupling of the up-type quarks to $\Phi_1$
\cite{2h-11}.

The tree-level couplings of the neutral Higgs bosons can have
sizable deviations from those of SM Higgs boson. Table \ref{dlcoup}
shows the couplings of neutral Higgs bosons with respect to the SM
Higgs boson. According to Table \ref{dlcoup}, the A2HDM can be
mapped to the four traditional types of 2HDMs via the angles
$\theta_d$ and $\theta_l$ specified in Table \ref{dltype}.

%%%%%%%%%%%%%%%%%%%%
\begin{table}
\caption{The tree-level couplings of the neutral Higgs bosons with
respect to those of the SM Higgs boson. $u$, $d$ and $l$ denote the
up-type quarks, down-type quarks and the charged leptons,
respectively. The angle $\alpha$ parameterizes the mixing of two
CP-even Higgses $h$ and $H$.} \vspace{0.5cm}
  \setlength{\tabcolsep}{2pt}
  \centering
  \begin{tabular}{|c|c|c|c|c|}
    \hline
     &~~$VV$~$(WW,~ZZ)$~~& ~~~~$u\bar{u}$~~~~ &~~~~ $d\bar{d}$~~~~&~~~~ $l\bar{l}$~~~~\\
    \hline
     $~h~$
     & $\sin(\beta-\alpha)$ & $\frac{\cos\alpha}{\sin\beta}$
     & $-\frac{\sin(\alpha-\theta_d)}{\cos(\beta-\theta_d)}$
     & $-\frac{\sin(\alpha-\theta_l)}{\cos(\beta-\theta_l)}$
     \\
     $~H~$
      & $\cos(\beta-\alpha)$ &$\frac{\sin\alpha}{\sin\beta}$
     &$\frac{\cos(\alpha-\theta_d)}{\cos(\beta-\theta_d)}$
     &$\frac{\cos(\alpha-\theta_l)}{\cos(\beta-\theta_l)}$
     \\
     $~A~$
     & 0 & $-\frac{i}{\tan\beta}\gamma_5$
     & $-i\tan(\beta-\theta_d)\gamma_5$
     & $-i\tan(\beta-\theta_l)\gamma_5$
     \\
     \hline

      \end{tabular}
\label{dlcoup}
\end{table}
%%%%%%%%%%%%%%%%%%%%%%%%%%%%%%%%%%%%%%%%%%%%%%%%%

\vspace{3.0cm}
%%%%%%%%%%%%%%%%%%%%
\begin{table}
 \caption{The values of mixing angles $\theta_d$ and
$\theta_l$ for the four traditional types of 2HDMs.}\vspace{0.5cm}
  \setlength{\tabcolsep}{2pt}
  \centering
  \begin{tabular}{|c|c|c|c|c|}
    \hline
     &~~Type~I~~& ~~~~Type~II~~~~ &~~Lepton-specific~~&~~~~ Flipped~~~~\\
    \hline
     ~~$\theta_d$~~
     & $\frac{\pi}{2}$ & $0$
     & $\frac{\pi}{2}$ & $0$
     \\
     ~~$\theta_l$~~
      & $\frac{\pi}{2}$ & $0$
      & $0$ & $\frac{\pi}{2}$
     \\
     \hline
      \end{tabular}
\label{dltype}
\end{table}
%%%%%%%%%%%%%%%%%%%%%%%%%%%%%%%%%%%%%%%%%%%%%%%%%

\section{numerical calculations}
We have employed the following four codes to implement the various
theoretical and experimental constraints. We require the A2HDM to
explain the experimental data of flavor observables and the
electroweak precision data within 2$\sigma$ range.
\begin{itemize}
\item\textsf{2HDMC-1.5} \cite{2hc-1}: The code is used to implement the
theoretical constraints from the vacuum stability, unitarity and
coupling-constant perturbativity. Also the oblique parameters ($S$,
$T$, $U$) and $\delta\rho$ are calculated and the corresponding
experimental data are from \cite{stupara}. $\delta\rho$ has been
measured very precisely via Z-pole precision observables to be very
close to 1, which imposes a strong constraint on the mass difference
between the various Higgses in 2HDMs. In addition, the code
\textsf{2HDMC-1.5} \footnote{A bug is modified: $\Gamma(h\to
Z\gamma)=0$ for $m_h<m_Z$.} which calculates the Higgs couplings and
the decay branching fractions, provides the necessary inputs for the
following three codes.

\item\textsf{SuperIso-3.3} \cite{spriso}: The code is used to implement the
constraints from flavor observables, including $B\to X_s\gamma$
\cite{bsrdstv}, $B_s\to\mu^+\mu^-$ \cite{bsuu}, $B_u\to\tau\nu$
\cite{butv} and $D_s\to\tau\nu$ \cite{bsrdstv}. Also the constrains
from $\Delta m_{B_d}$ \cite{deltamdms} and $\Delta m_{B_s}$
\cite{deltamdms} are considered, which are calculated using the
formulas in \cite{deltmq}.

\item\textsf{HiggsBounds-4.1.0} \cite{hb-1,hb-2}: The code is used to implement
the exclusion constraints from the neutral and charged Higgses
searches at LEP, Tevatron and LHC at 95\% confidence level.

\item\textsf{HiggsSignals-1.1.0} \cite{hs-1,hs-2}: The code is used to perform a
global $\chi^2$ fit to the most up-to-date signal strength
measurements as of November 2013. We consider the 73 Higgs signal
strengths observables from ATLAS
\cite{alt-1,alt-2,alt-3,alt-4,alt-5,alt-6,alt-7,alt-8,alt-9}, CMS
\cite{cms-1,cms-2,cms-3,cms-4,cms-5,cms-6,cms-7,cms-8,cms-9,cms-10,cms-11,cms-12,cms-13},
CDF \cite{cdf-1} and D0 \cite{d0-1} collaborations as well as the
four Higgs mass measurements from the ATLAS and CMS
$h\to\gamma\gamma$ and $h\to ZZ^*\to 4l$ analyses, which are listed
in the \cite{hs-2}. In our discussions, we will pay particular
attention to the surviving samples with $\chi^2-\chi^2_{\rm min}
\leq 6.18$, where $\chi^2_{\rm min}$ denotes the minimum $\chi^2$.
These samples correspond to the 95\% confidence level regions in any
two dimensional plane of the model parameters when explaining the
Higgs data (corresponding to be within $2\sigma$ range).

\end{itemize}

In our calculations, the inputs parameters are taken as $m^2_{12}$,
the physical Higgs masses ($m_h$, $m_H$, $m_A$, $m_{H^{\pm}}$ ), the
vacuum expectation value ratio ($\tan\beta$), the CP-even Higgs
mixing angle ($\alpha$), and the mixing angles of the down-type
quark and charge lepton Yukawa couplings ($\theta_d$, $\theta_l$).
We fix respectively $m_h$ and $m_H$ as 125.5 GeV, and scan randomly
the parameters in the following ranges:
\begin{eqnarray}
&&50 {\rm\  GeV} \leq m_A,~m_{H^\pm}  \leq 900  {\rm\  GeV},\nn\\
&&-1 \leq \sin(\beta-\alpha) \leq 1, \hspace{1.6cm} 0.1 \leq \tan \beta \leq 50,\nn\\
&&0\leq \theta_d \leq \pi, \hspace{3.4cm} 0\leq \theta_l \leq \pi,\nn\\
&&m^2_{12}\ ({\rm
GeV^2})=\pm(0.1)^2,~\pm(1)^2,~\pm(5)^2,~\pm(10)^2,~\pm(30)^2,~\pm(50)^2,\nn\\
&&\hspace{2.7cm}\pm(100)^2,~\pm(180)^2,~\pm(300)^2,~\pm(400)^2,~\pm(500)^2,\nn\\
&&{\rm Scenario~A}:~ m_h=125.5~{\rm GeV}, ~~~~ 125.5~{\rm GeV }\leq m_H\leq900~{\rm GeV},\nn\\
&&{\rm Scenario~B}: m_H=125.5~{\rm GeV},~~~~  20~{\rm GeV }\leq
m_h\leq125.5~{\rm GeV}.
\end{eqnarray}

$\textsf{HiggsSignals-1.1.0}$ automatically consider the effects of
any neutral Higgs boson on $\chi^2$ if its mass satisfies \beq
|m_{h_i}-\hat{m}_s|~\leq~\Delta \hat{m}_s.\label{massreso}\eeq Where
$h_i$ denotes $h$, $H$ and $A$. $\hat{m}_s$ is the mass of signal
$s$ and $\Delta \hat{m}_s$ is the experimental mass resolution of
the analysis associated to signal $s$. However, if the $\chi^2$
contribution from the measured Higgs mass is activated, the
combinations with a Higgs boson mass which does not fulfill Eq.
(\ref{massreso}) are still considered. For the detailed introduction
on the calculation of $\chi^2$, see \cite{hs-1,hs-2}.

\section{results and discussions}
\subsection{Scenario A}

%%%%%%%%%%%%%%%%%%%%%
\begin{figure}[tb]
%\begin{center}
 \epsfig{file=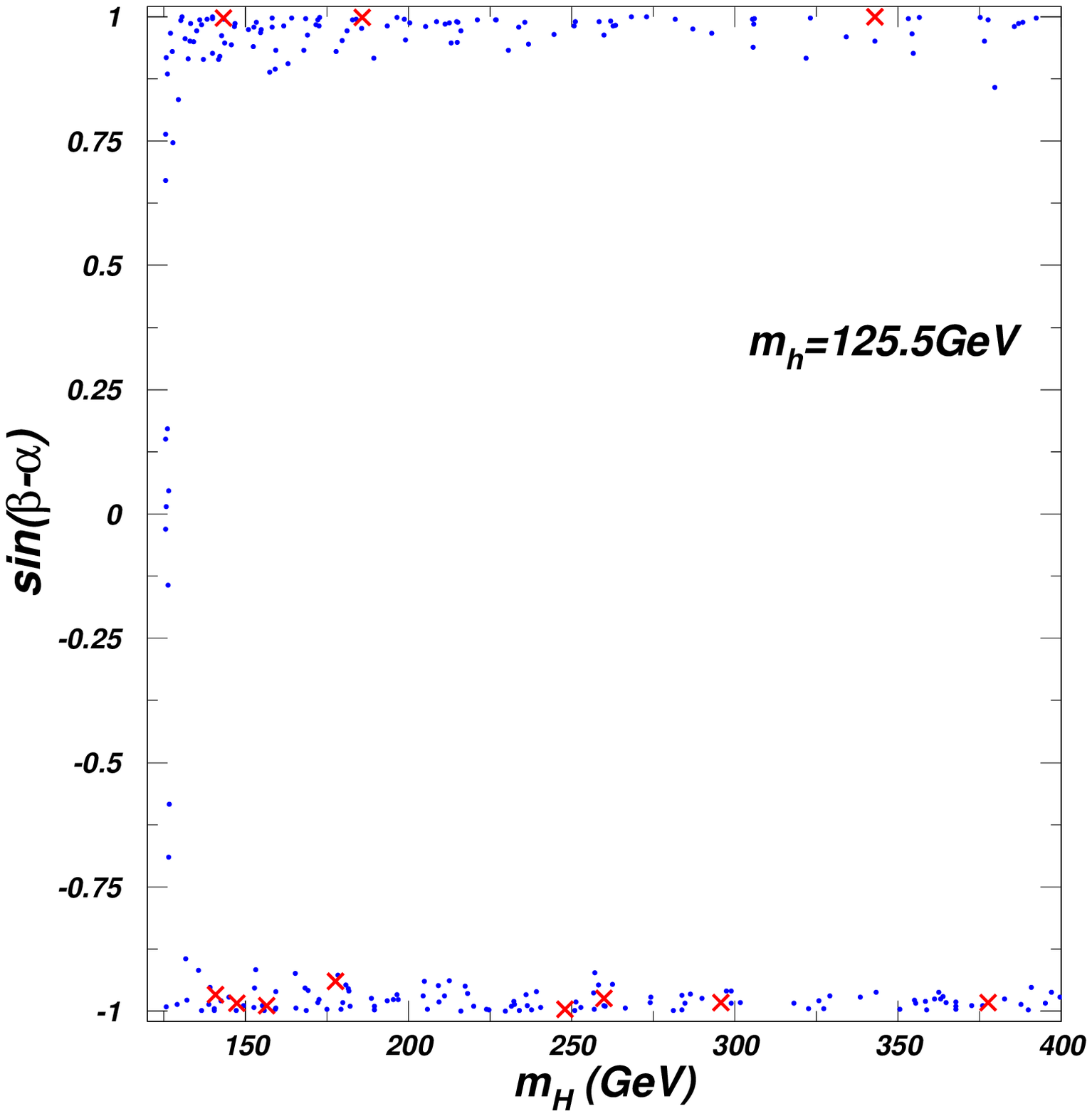,height=8.32cm}
  \epsfig{file=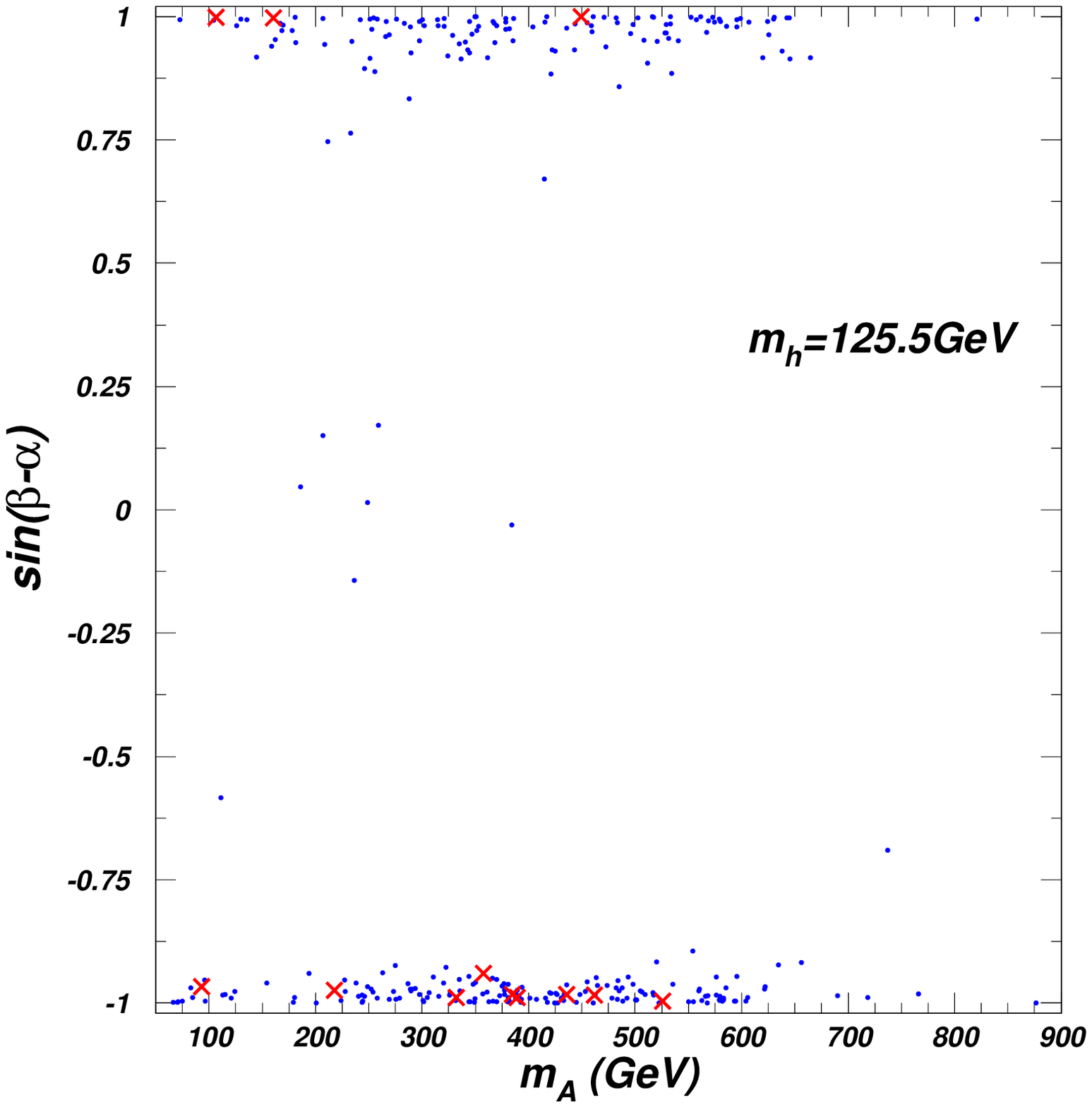,height=8.32cm}
%\end{center}
\vspace{-0.2cm} \caption{The scatter plots of surviving samples in
scenario A projected on the planes of $\sin(\beta-\alpha)$ versus
$m_H$ and $\sin(\beta-\alpha)$ versus $m_A$. The crosses (red), and
bullets (blue) samples respectively have the values of $\chi^2$ in
the ranges of $81.0\thicksim82.2$ and $82.2\thicksim87.2$, where the
three values are respectively the minimal value of $\chi^2$ in
scenario A ($\chi^2_{Amin}$), the SM value ($\chi^2_{SM}$) and the
value of $\chi^2$ at 2$\sigma$ level in scenario A
($\chi^2_{A2\sigma}$).} \label{a-sba-h}
\end{figure}
%%%%%%%%%%%%%%%%%%%%

%%%%%%%%%%%%%%%%%%%%%
\begin{figure}[tb]
%\begin{center}
 \epsfig{file=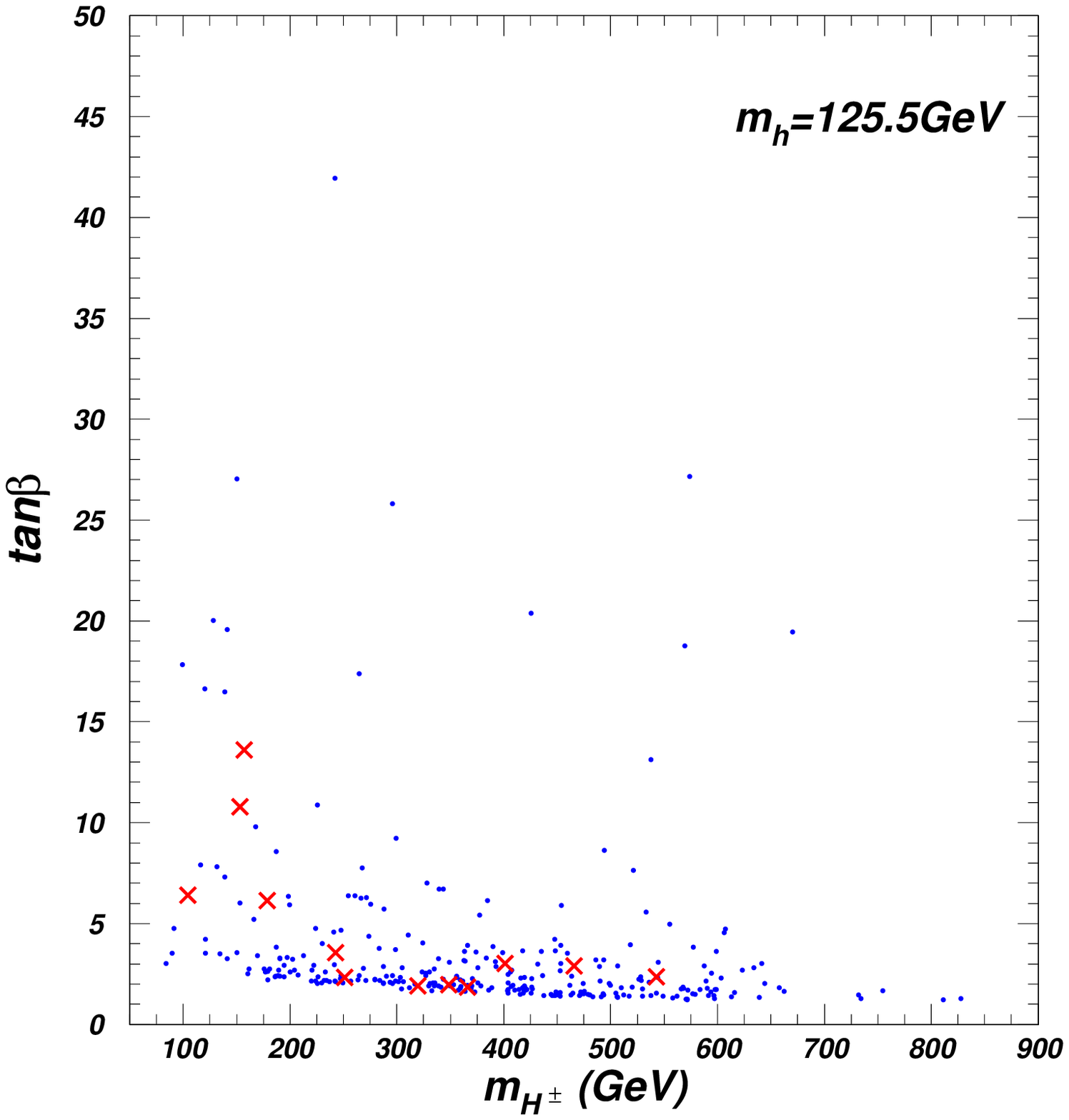,height=8.32cm}
  \epsfig{file=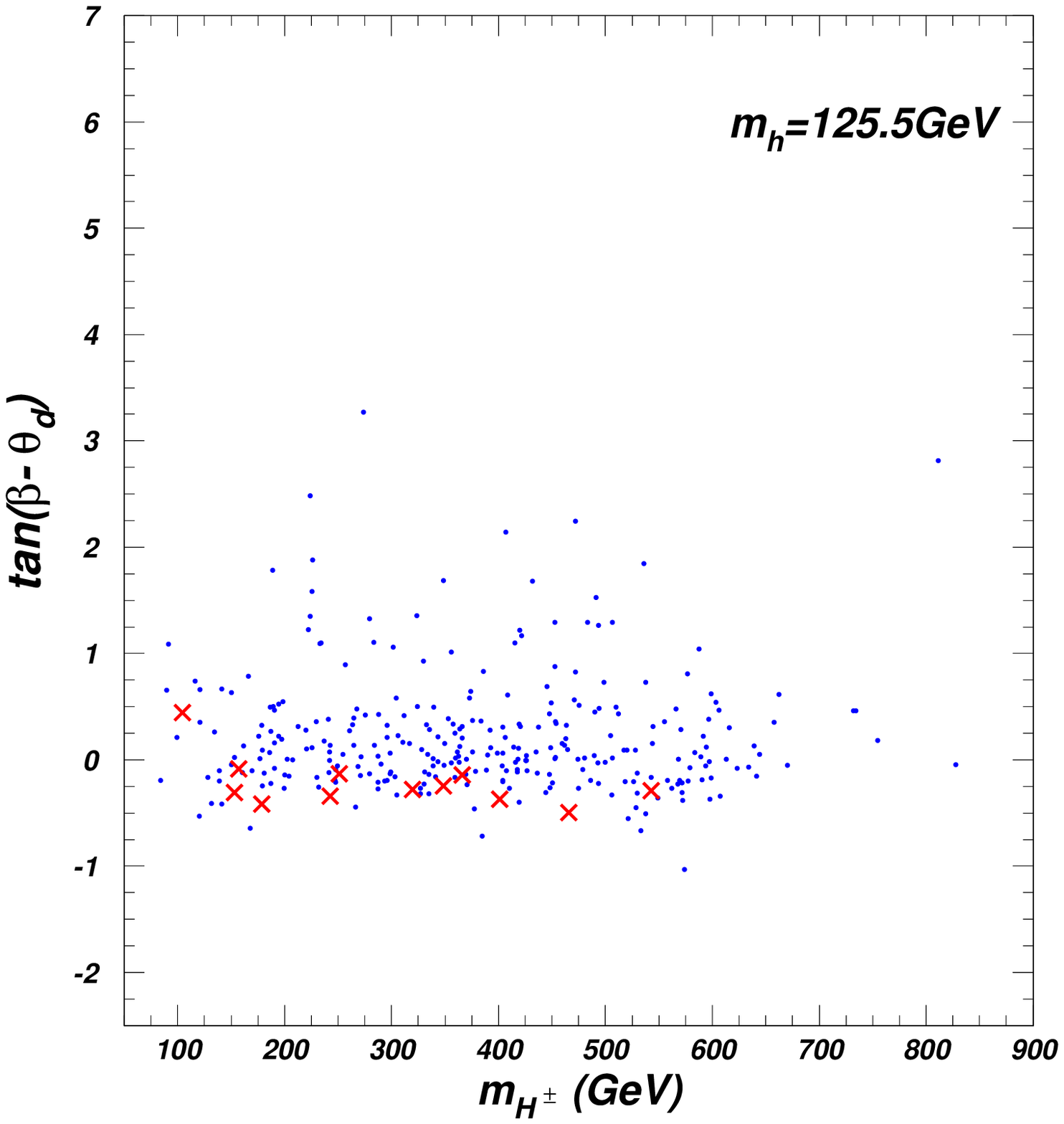,height=8.32cm}
%\end{center}
\vspace{-0.4cm} \caption{Same as Fig. \ref{a-sba-h}, but projected
on the planes of $\tan\beta$ versus $m_{H^{\pm}}$ and
$\tan(\beta-\theta_d)$ versus $m_{H^{\pm}}$. } \label{a-tb-hp}
\end{figure}
%%%%%%%%%%%%%%%%%%%%

%%%%%%%%%%%%%%%%%%%%%
\begin{figure}[tb]
%\begin{center}
 \epsfig{file=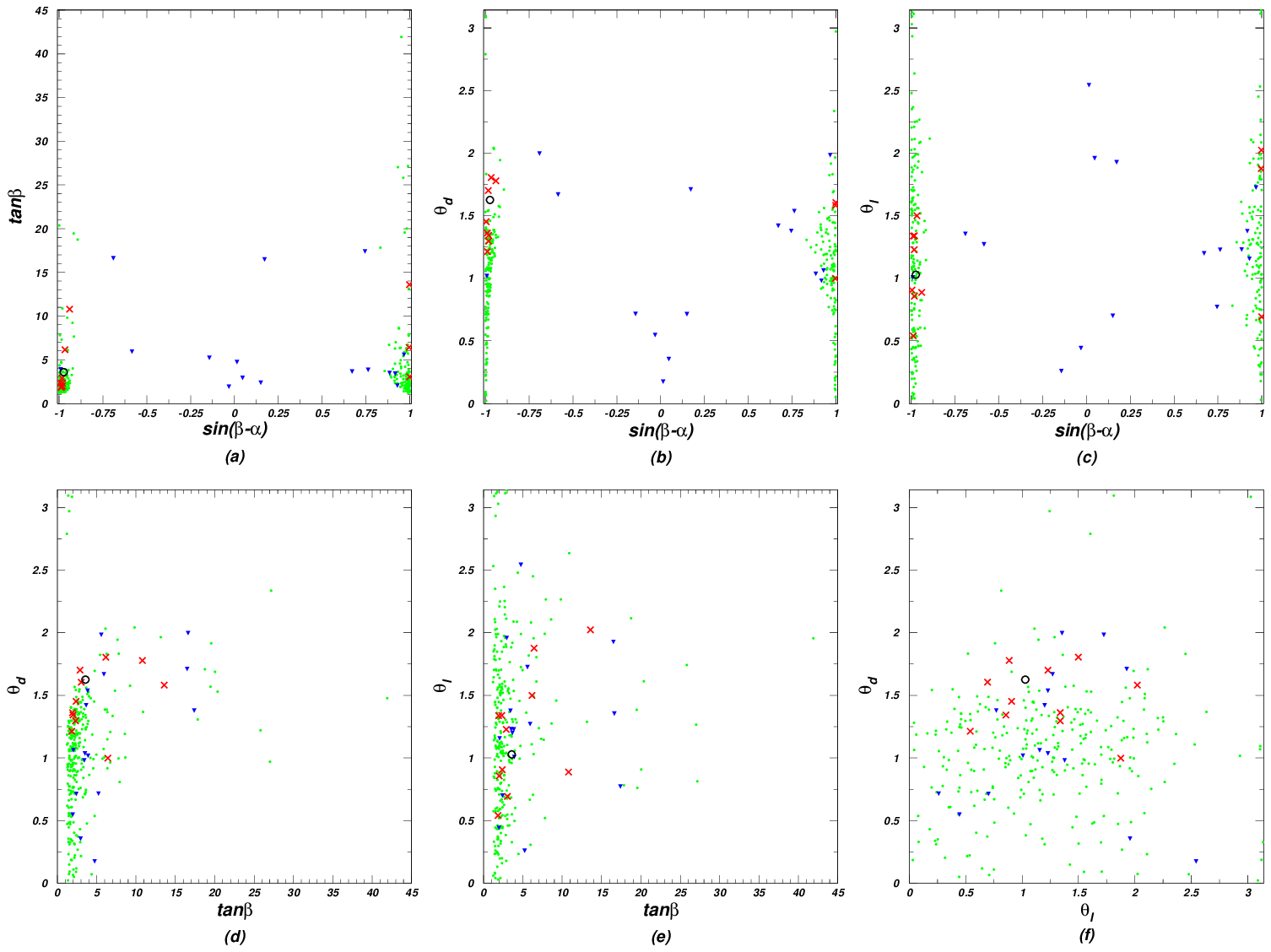,height=14.5cm,width=17.0cm}
%\end{center}
\vspace{-1.07cm} \caption{The scatter plots of surviving samples in
scenario A projected on the planes of mixing angles. The $\chi^2$
values of the crosses (red), bullets (green) and inverted triangles
(blue) samples are respectively in the ranges of
$\chi^2_{Amin}\thicksim\chi^2_{SM}$ and
$\chi^2_{SM}\thicksim\chi^2_{A2\sigma}$ for 128 GeV $\leq m_H\leq$
900 GeV, and $\chi^2_{SM}\thicksim\chi^2_{A2\sigma}$ for 125.5 GeV
$\leq m_H<$ 128 GeV. The $\chi^2$ values of the circle (black) is
$\chi^2_{Amin}$.} \label{a-mixing}
\end{figure}
%%%%%%%%%%%%%%%%%%%%

%%%%%%%%%%%%%%%%%%%%%
\begin{figure}[tb]
%\begin{center}
 \epsfig{file=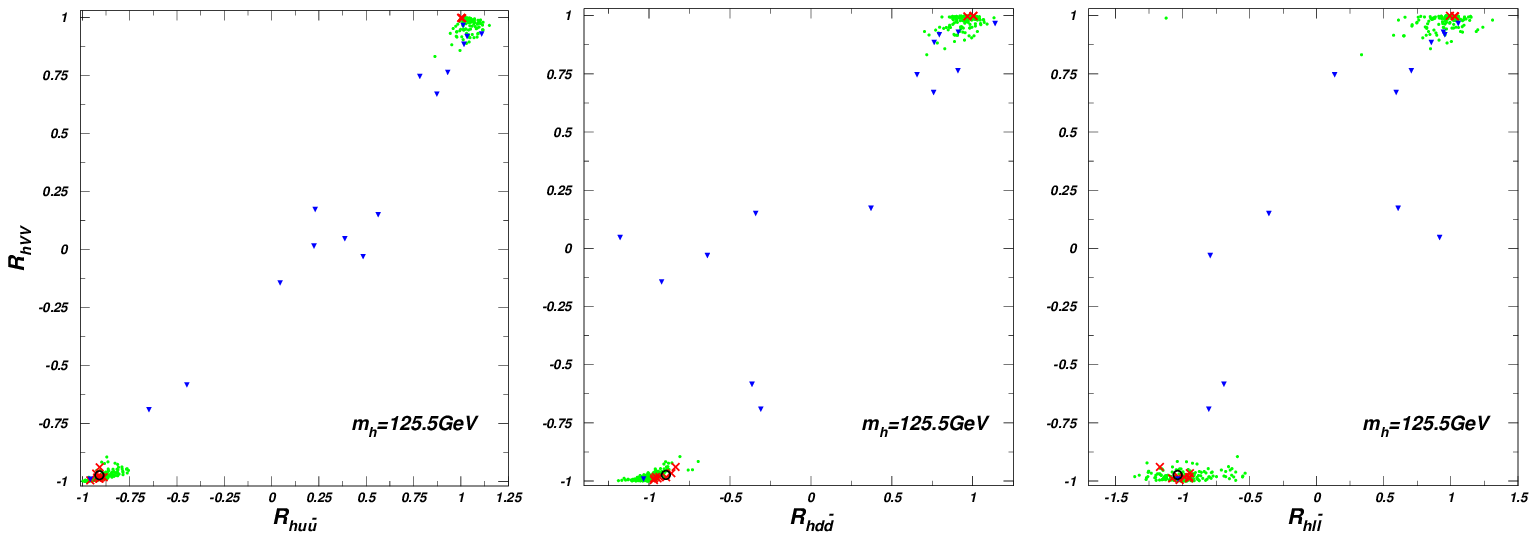,height=7.5cm,width=17cm}
%\end{center}
\vspace{-1.0cm} \caption{Same as Fig. \ref{a-mixing}, but projected
on the planes of $R_{hVV}$ versus $R_{hu\bar{u}}$, $R_{hVV}$ versus
$R_{hd\bar{d}}$ and $R_{hVV}$ versus $R_{hl\bar{l}}$. $R_{hVV}$ and
$R_{hf\bar{f}}$ denote the light CP-even Higgs couplings to gauge
bosons and $f\bar{f}$ ($f=u,~d,~l$) normalized to the SM couplings,
respectively.} \label{a-coup}
\end{figure}
%%%%%%%%%%%%%%%%%%%%

Let us begin by discussing the scenario A in which the mass of the
light CP-even Higgs $h$ is fixed as 125.5 GeV. In Fig.
\ref{a-sba-h}, we project the surviving samples with $\chi^2$ being
within $2\sigma$ range on the planes of $\sin(\beta-\alpha)$ versus
$m_H$ and $\sin(\beta-\alpha)$ versus $m_A$, respectively. The left
panel shows that, for the heavy CP-even Higgs mass is close to 125.5
GeV, it can give the important contributions to $\chi^2$, and the
absolute values of $\sin(\beta-\alpha)$ can be allowed to be as low
as 0, in which the $HVV$ couplings approach to SM while $hVV$
approach to 0. For $m_H\geq$ 128 GeV, $\sin(\beta-\alpha)$ is
allowed to be in the ranges of $0.83\thicksim1$ and
$-1\thicksim-0.89$. A small value of $\chi^2$ favors a large
absolute value of $\sin(\beta-\alpha)$, which denotes that the
absolute values of $hVV$ couplings approach to SM.

Unlike the heavy CP-even Higgs, the right panel of Fig.
\ref{a-sba-h} shows that the CP-odd Higgs $A$ does not give the very
visible effects on $\chi^2$ around 125.5 GeV compared to the other
mass ranges. $m_A$ is required to be larger than 63 GeV, and the
on-shell decay $h\to AA$ is kinematically forbidden, which hardly
affects the observed Higgs signals.

In Fig. \ref{a-tb-hp}, the surviving samples are projected on the
planes of $\tan\beta$ versus $m_{H^{\pm}}$ and $\tan(\beta-\alpha)$
versus $m_{H^{\pm}}$. The left panel shows that the surviving
samples favor $1<\tan\beta<5$ and allow $\tan\beta>$ 30 for
$m_{H^{\pm}}>$ 230 GeV. The constraints from $\Delta m_{B_d}$ and
$\Delta m_{B_s}$ require $\tan\beta$ to be
 larger than 1 for the whole range of $m_{H^{\pm}}$, and larger than 3 for $m_{H^{\pm}}<100$ GeV.
The right panel shows that the surviving samples favor -0.5
$<\tan(\beta-\theta_d)<$ 0.5. The flavor interactions mediated by
$H^{\pm}$ are proportional to $\tan(\beta-\theta_d)$. The
constraints from the flavor observables allow $m_{H^{\pm}}$ to be
smaller than 100 GeV for the very small absolute of
$\tan(\beta-\theta_d)$, and $\tan(\beta-\theta_d)$ to be larger than
3 for $m_{H^{\pm}}>$ 250 GeV. In addition, the samples with smaller
$\chi^2$ than SM favor $\tan(\beta-\theta_d)$ to be in the range of
$-0.5\thicksim0$ for $m_{H^{\pm}}>150$ GeV.

The contributions of the heavy CP-even Higgs boson to $\chi^2$ can
be sizably suppressed for $m_H\geq$ 128 GeV. Therefore, we classify
the surviving samples into groups: 125.5 GeV $\leq m_H<$ 128 GeV and
128 GeV $\leq m_H\leq$ 900 GeV. In Fig. \ref{a-mixing}, the two
groups of surviving samples are projected on the planes of mixing
angles ($\sin(\beta-\alpha)$, $\tan\beta$, $\theta_d$ and
$\theta_l$). Fig. \ref{a-mixing} (a) shows that $\tan\beta$ can be
over 20 for $\sin(\beta-\alpha)$ is close to 1. Fig. \ref{a-mixing}
(b) shows that, for $m_H>$ 128 GeV, the mixing angle $\theta_d$ can
loose constraints on $\sin(\beta-\alpha)$ visibly. For example, for
$\theta_d\simeq0$ (Type-II and Flipped 2HDMs), the absolute value of
$\sin(\beta-\alpha)$ is required to be very close to 1. While
$\sin(\beta-\alpha)$ are allowed to vary in the range of
$0.83\thicksim1$ and $-1\thicksim-0.89$ for $\theta_d$ has the
properly large value. Also Fig. \ref{a-mixing} (c) shows that
$\sin(\beta-\alpha)$ in the positive range is required to be very
close to 1 for $\theta_l\simeq0$ (Type-II and Lepton-specific
2HDMs).

According to Figs. \ref{a-mixing} (d) and (e), although the
surviving samples favor a small value of $\tan\beta$, the value of
$\chi^2$ can be smaller than SM for a large $\tan\beta$ when
$\theta_d$ and $\theta_l$ have the proper large values, such as
$\tan\beta=$13.5, $\theta_d=1.6$ and $\theta_l=2.0$.

Fig. \ref{a-mixing} (f) shows that the samples with smaller $\chi^2$
than SM are favored in the range of 1 $<\theta_d<$ 2 and 0.5
$<\theta_l<$ 2.2. Thus, it is possible that Type-I 2HDM gives the
smaller value of $\chi^2$ than SM. The minimal value of $\chi^2$
(81.0) appears at $\theta_d=$ 1.7 and $\theta_l=$ 1.3.

In Fig. \ref{a-coup}, the surviving samples are projected on the
planes of Higgs couplings. For 125.5 GeV $\leq m_H<$ 128 GeV, the
heavy CP-even Higgs gives the important contributions to $\chi^2$.
Therefore, there may be sizable deviations from SM for the couplings
$hVV$, $hu\bar{u}$, $hd\bar{d}$ and $hl\bar{l}$. For $m_H\geq$ 128
GeV and the $hVV$ coupling with the small absolute value, the
$hb\bar{b}$ coupling by suppressed properly is required to obtain
enough large $Br(h\to ZZ^*)$ and $Br(h\to \gamma\gamma)$. The
$h\to\gamma\gamma$ and $h\to ZZ^*\to 4l$ have the rather precise
measurements and mass resolution, which play a very important role
in the calculations of $\chi^2$. The signal strengths of $h\to
\tau\tau$ have a large uncertainty and the signals are not important
in the calculations of $\chi^2$. In addition, the mass resolution of
$h\to \tau\tau$ is 20 GeV for the analysis of ATLAS
\cite{alt-6,alt-7} and 25 GeV for CMS \cite{cms-9}, CDF \cite{cdf-1}
and D0 \cite{d0-1}. Therefore, $H$ and $A$ with $100\thicksim150$
GeV may contribute to $\chi^2$. The constraints on $h\tau\bar{\tau}$
is much more weaken than $hu\bar{u}$ and $hd\bar{d}$

For the samples with smaller $\chi^2$ than SM, there is the same
sign for the light CP-even Higgs couplings to fermions and gauge
bosons. Compared to SM, the $hVV$, $hu\bar{u}$ and $hd\bar{d}$
couplings are suppressed, and the suppressions are allowed to be as
low as 0.94, 0.90 and 0.83, while the absolute value of
$R_{hl\bar{l}}$ are allowed to be as high as 1.2.

\subsection{Scenario B}

%%%%%%%%%%%%%%%%%%%%%
\begin{figure}[tb]
%\begin{center}
 \epsfig{file=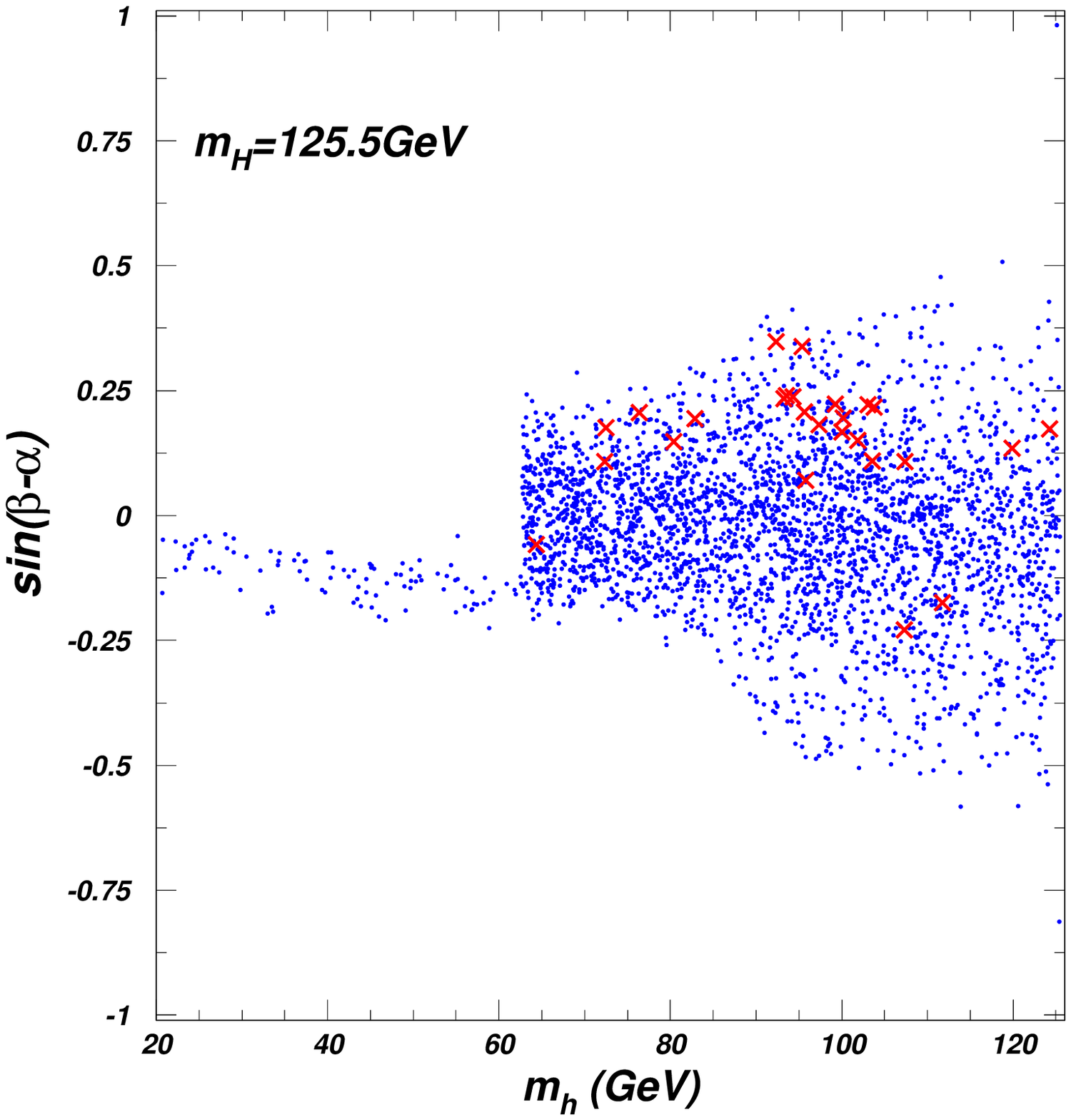,height=8.32cm}
  \epsfig{file=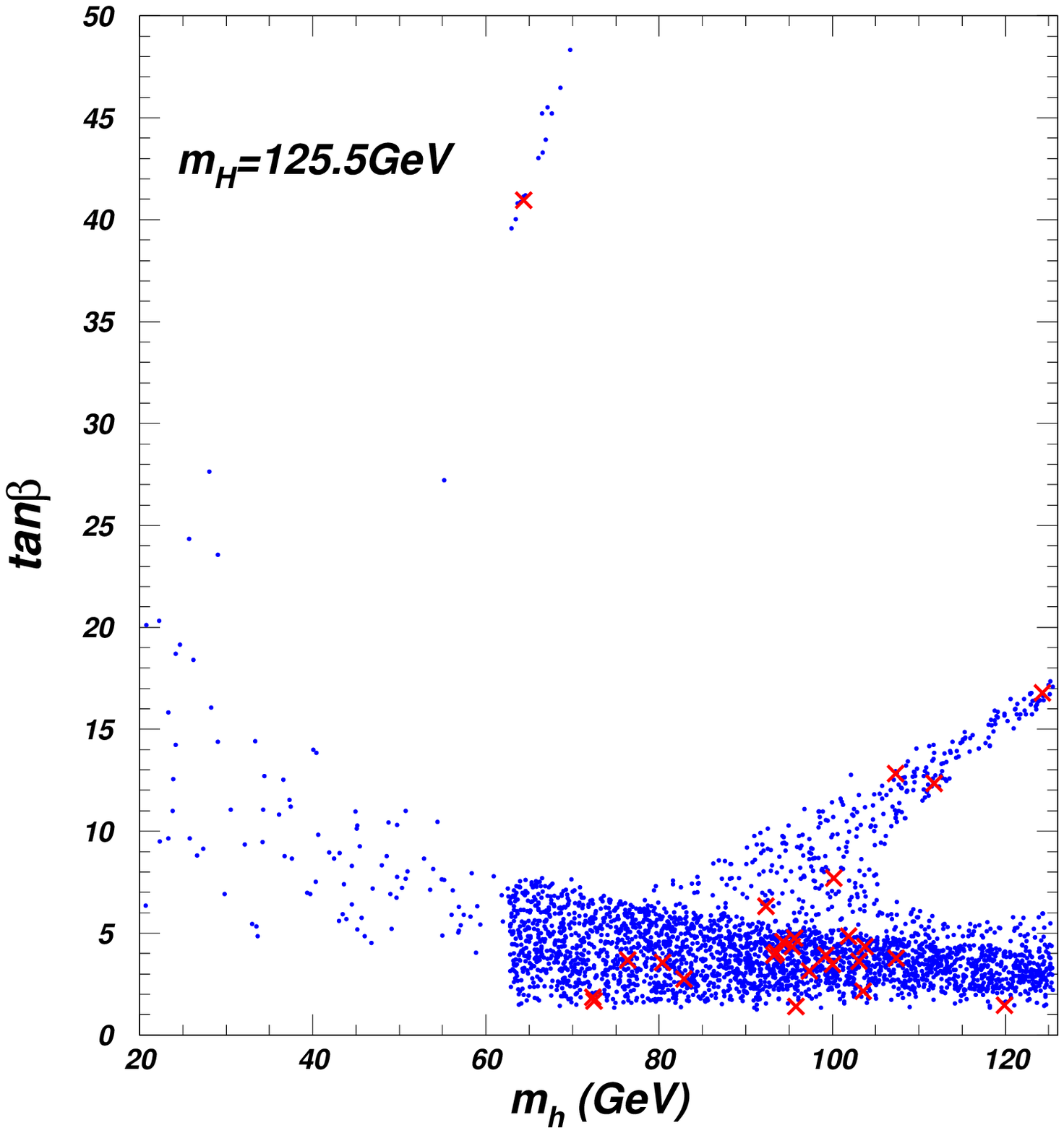,height=8.32cm}
%\end{center}
\vspace{-0.4cm} \caption{The scatter plots of surviving samples in
scenario B projected on the planes of $\sin(\beta-\alpha)$ versus
$m_h$ and $\tan\beta$ versus $m_h$. The crosses (red) and bullets
(blue) samples respectively have the values of $\chi^2$ in the
ranges of $81.5\thicksim82.2$ and $82.2\thicksim87.7$, where the
three values are respectively the minimal value of $\chi^2$ in
scenario B ($\chi^2_{Bmin}$), the SM value ($\chi^2_{SM}$) and the
value of $\chi^2$ at 2$\sigma$ level in scenario B
($\chi^2_{B2\sigma}$).} \label{b-sba-h}
\end{figure}
%%%%%%%%%%%%%%%%%%%%

%%%%%%%%%%%%%%%%%%%%%
\begin{figure}[tb]
%\begin{center}
 \epsfig{file=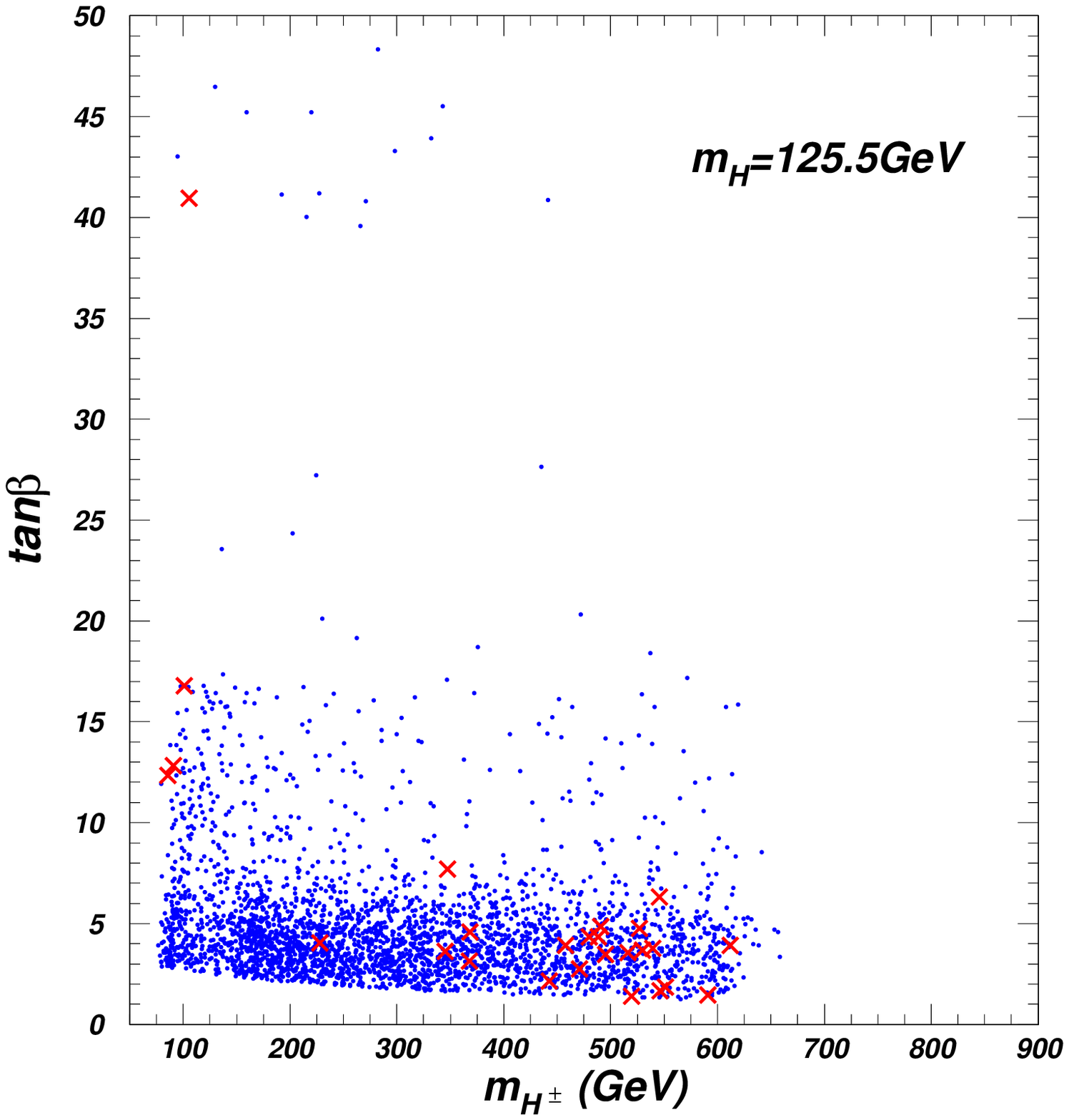,height=8.32cm}
  \epsfig{file=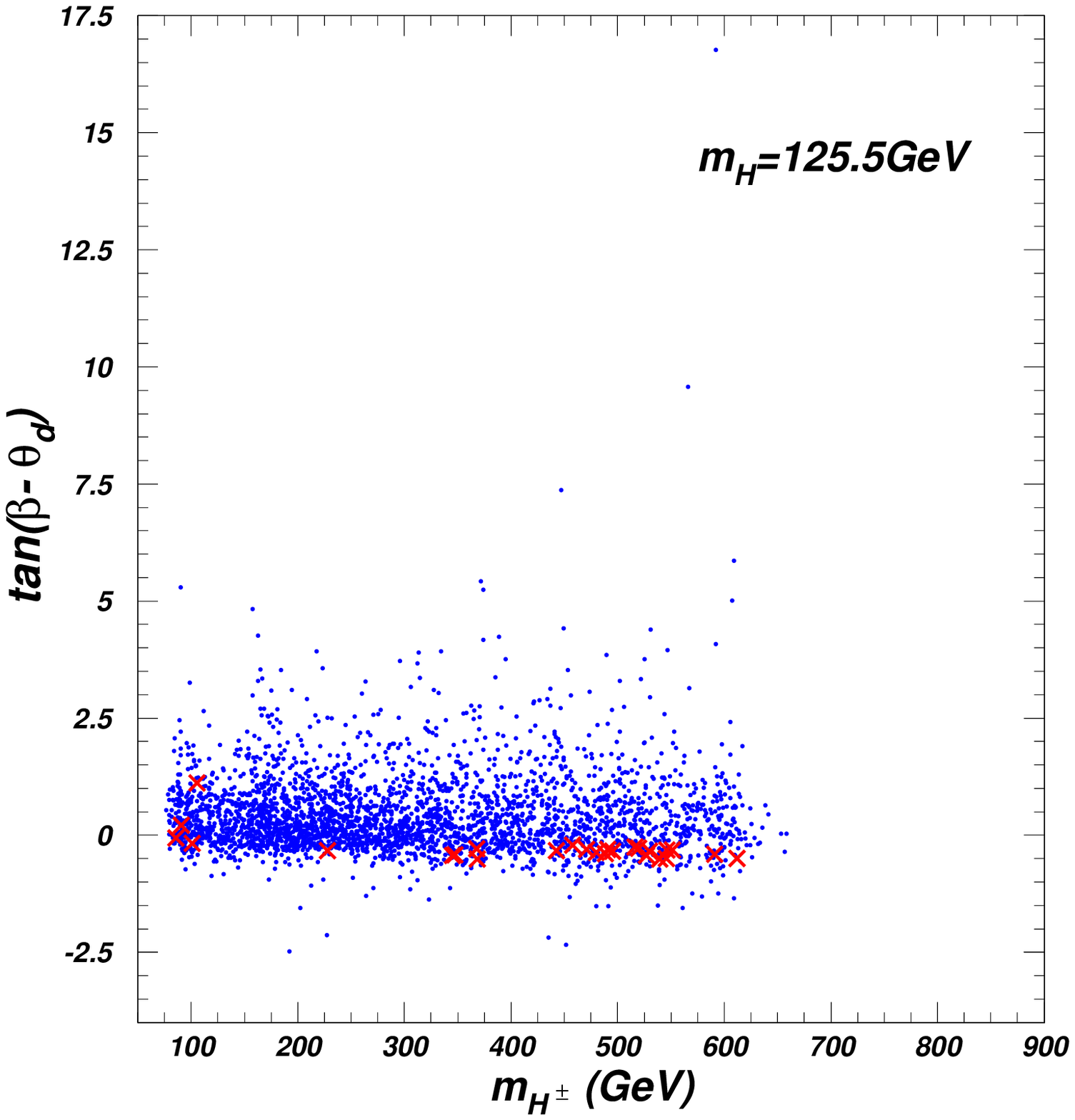,height=8.32cm}
%\end{center}
\vspace{-0.4cm} \caption{Same as Fig. \ref{b-sba-h}, but projected
on the planes of $\tan\beta$ versus $m_{H^{\pm}}$ and
$\tan(\beta-\theta_d)$ versus $m_{H^{\pm}}$.} \label{b-tb-hp}
\end{figure}
%%%%%%%%%%%%%%%%%%%%

Now we discuss the scenario B in which the mass of the heavy CP-even
Higgs $H$ is fixed as 125.5 GeV. In Fig. \ref{b-sba-h}, we project
the surviving samples with $\chi^2$ being within $2\sigma$ range on
the planes of $\sin(\beta-\alpha)$ versus $m_h$ and $\tan\beta$
versus $m_h$, respectively. The left panel shows that,  for 125 GeV
$\leq m_h\leq$ 125.5 GeV, the absolute values of
$\sin(\beta-\alpha)$ can be allowed to be as high as 1, which
denotes $hVV$ couplings approach to SM while $HVV$ approach to 0.
Such light CP-even Higgs can give the important contributions to
$\chi^2$. The minimal absolute value of $\sin(\beta-\alpha)$
decreases with $m_h$ in principle. The light CP-even Higgs can be
allowed to be as low as 20 GeV for -0.25 $<\sin(\beta-\alpha)\leq$
0. To be consistent with LEP constraints, the suppression of
$hb\bar{b}$ coupling is also required for some surviving samples in
addition to the small absolute value of $\sin(\beta-\alpha)$. In
addition, the small values of $\chi^2$ favor -0.25
$<\sin(\beta-\alpha)<$ 0.38, which denotes that the absolute values
of $HVV$ couplings are close to SM. The right panel shows that
$\tan\beta$ is required to be larger than 4 for $m_h<$ 60 GeV, which
is due to the constraints of the observed Higgs signals on the
opening decay $H\to hh$.

In Fig. \ref{b-tb-hp}, the surviving samples are projected on the
planes of $\tan\beta$ versus $m_{H^{\pm}}$ and
$\tan(\beta-\theta_d)$ versus $m_{H^{\pm}}$. The left panel shows
that the surviving samples favor 1 $<\tan\beta<$ 7 and allow
$\tan\beta>$ 40 for the proper $m_{H^{\pm}}$. Similar to scenario A,
$\tan\beta$ is required to be larger than 1 for the whole range of
$m_{H^{\pm}}$, and larger than 3 for the $m_{H^{\pm}}<100$ GeV. The
right panel shows that the surviving samples favor -1
$<\tan(\beta-\theta_d)<$ 2.5. The constraints from the flavor
observables require the absolute value of $\tan(\beta-\theta_d)$ to
be smaller than 2.5 for $m_{H^{\pm}}<100$ GeV, and allow
$\tan(\beta-\theta_d)$ to be larger than 10 for $m_{H^{\pm}}>600$
GeV. The samples with smaller $\chi^2$ than SM favor
$\tan(\beta-\theta_d)$ to be in the range of $-0.5\thicksim0$ for
the large $m_{H^{\pm}}$ and be enhanced for $m_{H^{\pm}}$ around 100
GeV.

%%%%%%%%%%%%%%%%%%%%%
\begin{figure}[tb]
%\begin{center}
 \epsfig{file=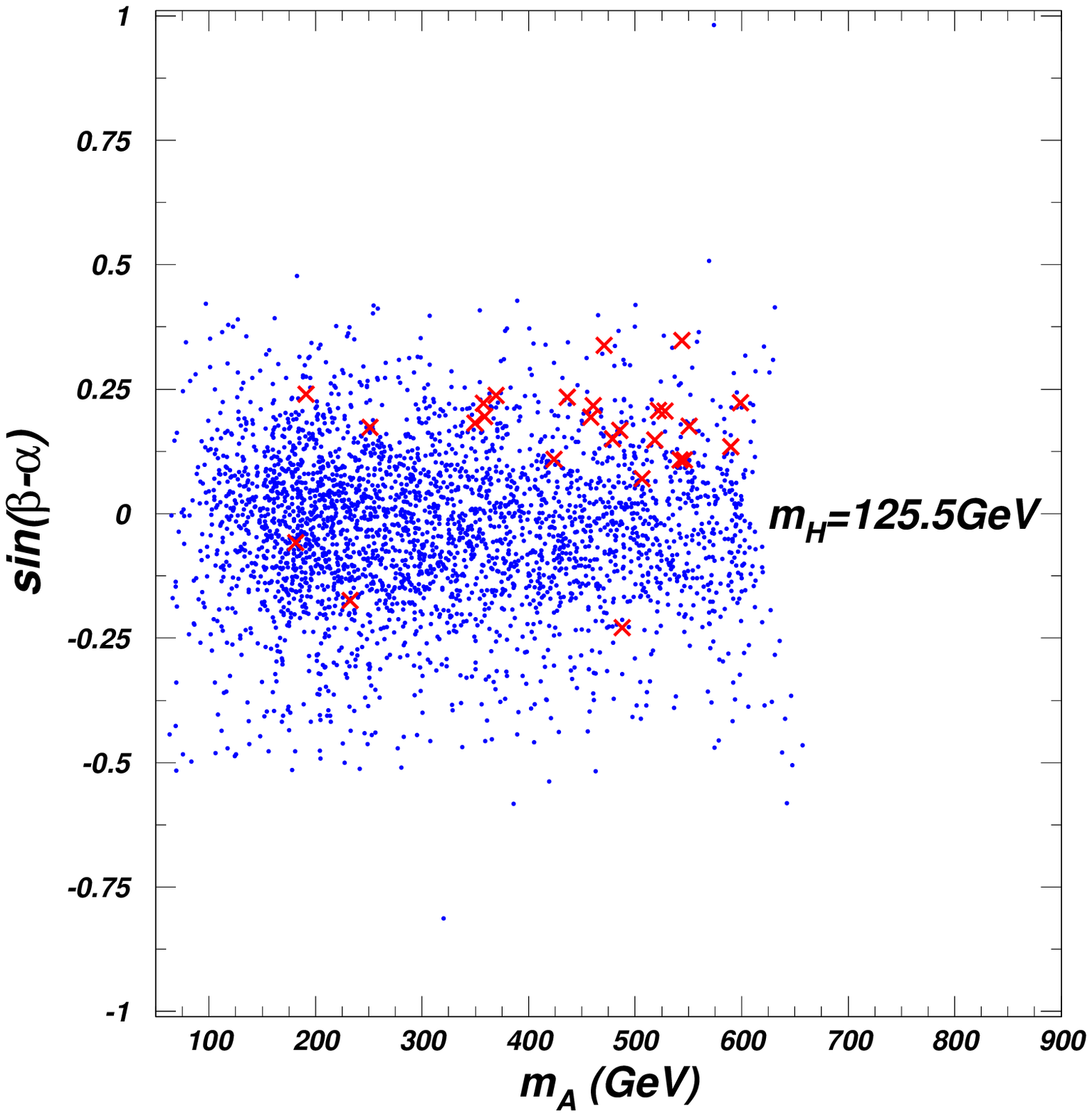,height=8.32cm}
  \epsfig{file=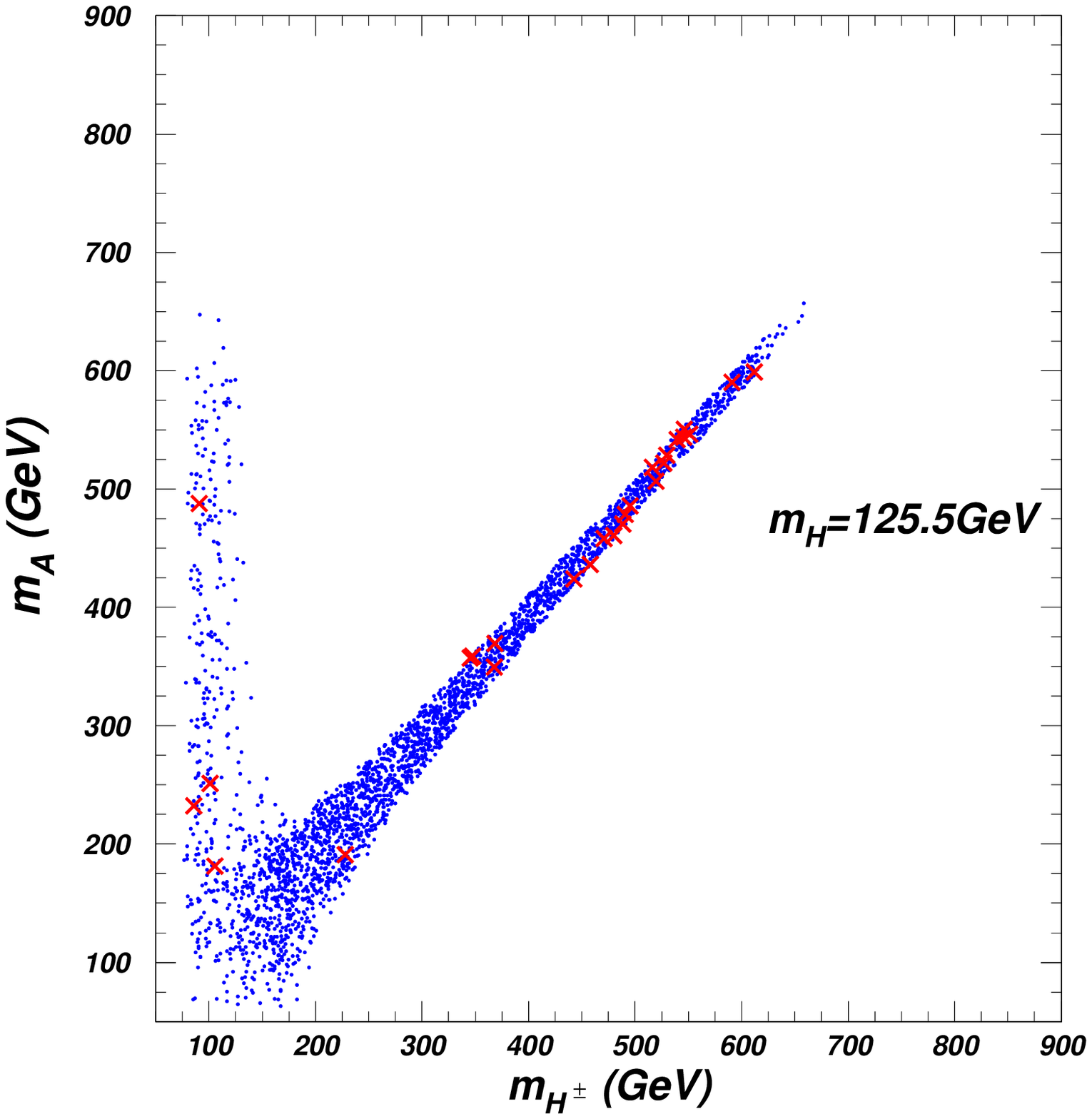,height=8.32cm}
%\end{center}
\vspace{-1.05cm} \caption{Same as Fig. \ref{b-sba-h}, but projected
on the planes of $\sin(\beta-\alpha)$ versus $m_A$ and $m_A$ versus
$m_{H^\pm}$.} \label{b-sba-a}
\end{figure}
%%%%%%%%%%%%%%%%%%%%

%%%%%%%%%%%%%%%%%%%%%
\begin{figure}[tb]
%\begin{center}
 \epsfig{file=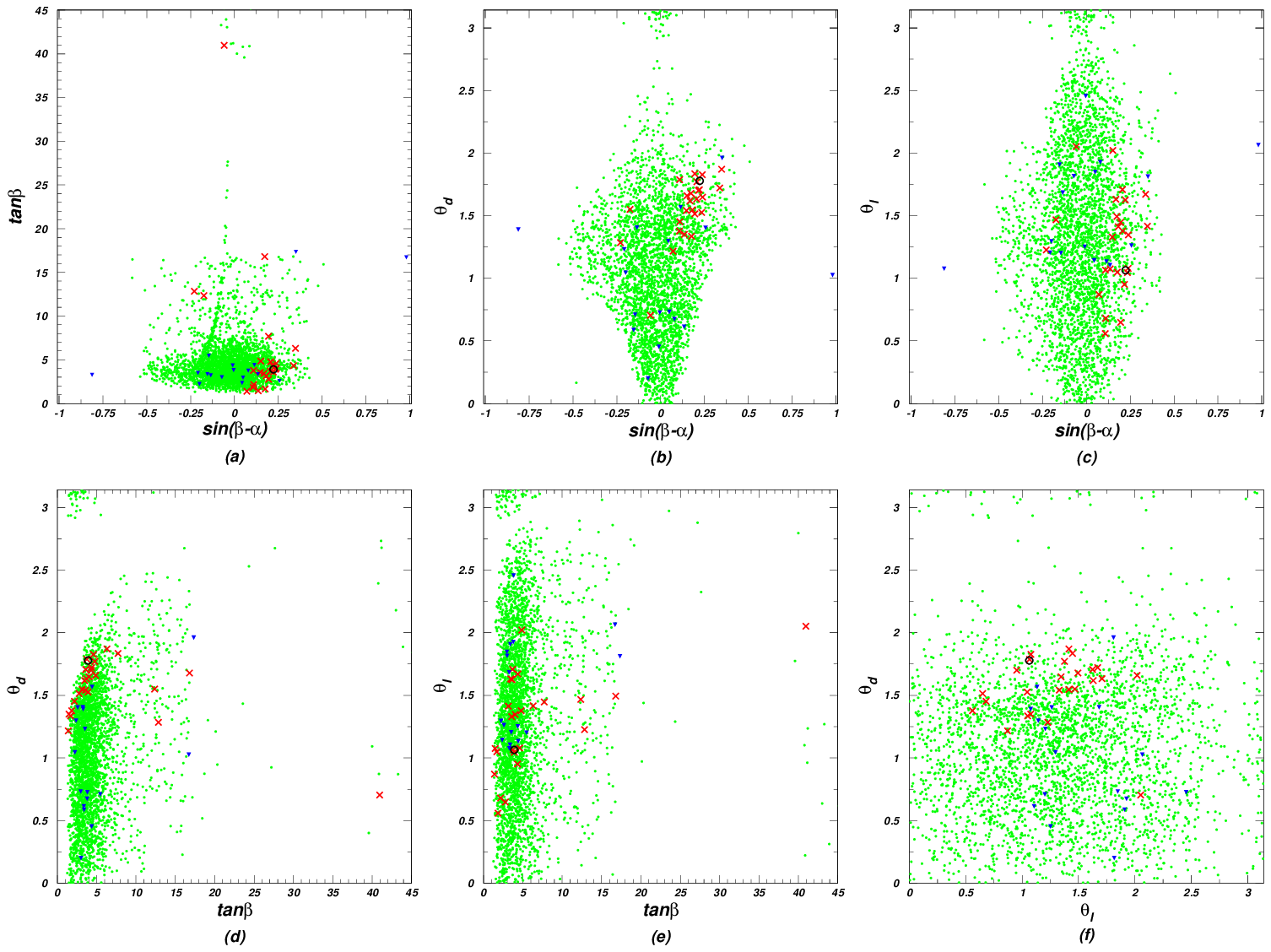,height=14.5cm,width=17.0cm}
%\end{center}
\vspace{-1.15cm} \caption{The scatter plots of surviving samples in
scenario B projected on the planes of mixing angles. The $\chi^2$
values of the crosses (red), bullets (green) and inverted triangles
(blue) samples are respectively in the ranges of
$\chi^2_{Bmin}\thicksim\chi^2_{SM}$ and
$\chi^2_{SM}\thicksim\chi^2_{B2\sigma}$ for 20 GeV $\leq m_H<$ 125
GeV, and $\chi^2_{SM}\thicksim\chi^2_{B2\sigma}$ for 125 GeV $\leq
m_H<$ 125.5 GeV. The $\chi^2$ values of the circle (black) is
$\chi^2_{Bmin}$.} \label{b-mixing}
\end{figure}
%%%%%%%%%%%%%%%%%%%%

%%%%%%%%%%%%%%%%%%%%%
\begin{figure}[tb]
%\begin{center}
 \epsfig{file=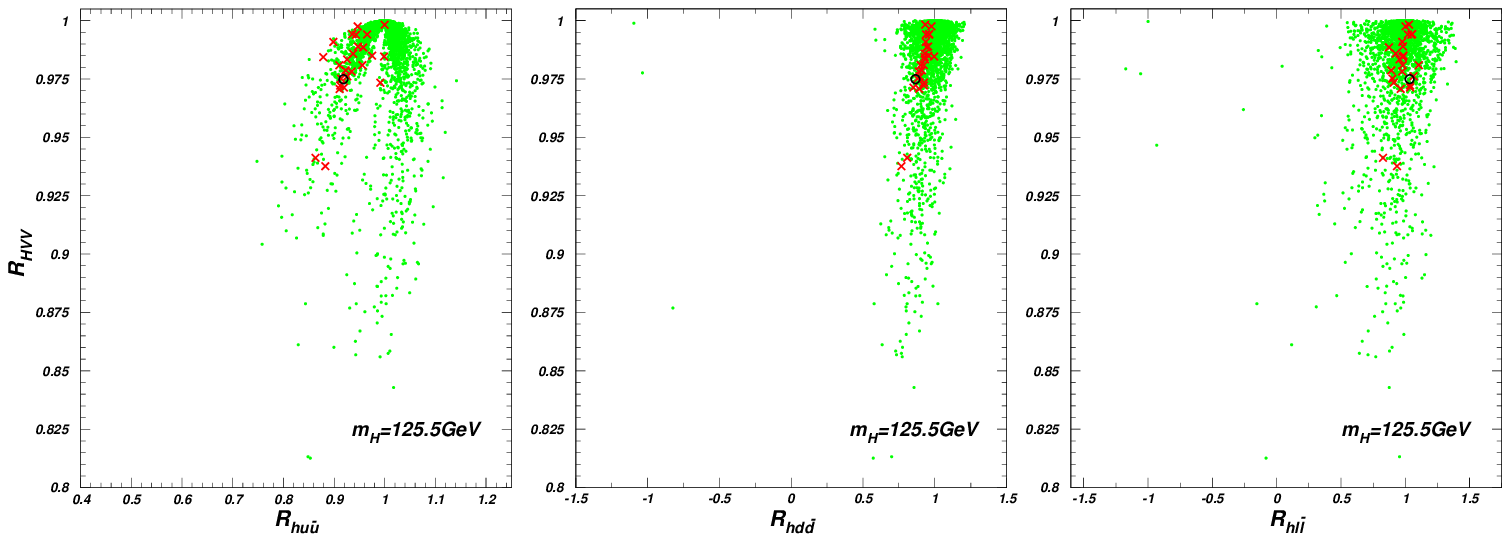,height=7.5cm,width=17cm}
%\end{center}
\vspace{-1.14cm} \caption{Same as Fig. \ref{b-mixing}, but only the
samples with 20 GeV $\leq m_h<$ 120 GeV projected on the planes of
$R_{HVV}$ versus $R_{Hu\bar{u}}$, $R_{HVV}$ versus $R_{Hd\bar{d}}$
and $R_{HVV}$ versus $R_{Hl\bar{l}}$. $R_{HVV}$ and $R_{Hf\bar{f}}$
denote the heavy CP-even Higgs couplings to gauge bosons and
$f\bar{f}$ ($f=u,~d,~l$) normalized to the SM couplings,
respectively.} \label{b-coup}
\end{figure}
%%%%%%%%%%%%%%%%%%%%
In Fig. \ref{b-sba-a}, the surviving samples are projected on the
planes of $\sin(\beta-\alpha)$ versus $m_{A}$ and $m_A$ versus
$m_{H^\pm}$. Similar to scenario A, the CP-odd Higgs $A$ does not
give the very visible effects on the $\chi^2$ around 125.5 GeV
compared to the other mass ranges. The on-shell decay $H\to AA$ is
kinematically forbidden, which hardly affects the observed Higgs
signals. The right panel shows that most of samples lie in the
region where there is small mass difference between $m_A$ and
$m_{H^{\pm}}$, and some other samples lie in the small region where
$m_{H^{\pm}}$ is around 100 GeV and has large mass difference from
$m_A$. Assuming $m_{12}^2=0$, Baradhwaj Coleppa et al. have shown
the strong correlations between $m_A$ and $m_{H^{\pm}}$ in the
Type-II 2HDM \cite{2h-3}. Here $m_{12}^2$ is taken as various
values, the strong correlations still exist but the latter region
becomes slightly wider than \cite{2h-3}. The main reason is from the
constraints of $\Delta\rho$, which is also studied in detail in
\cite{cus}. Since there is small mass difference between $m_h$ and
$m_H$ for the scenario B, $m_A$ and $m_{H^{\pm}}$ should have the
small mass difference to cancel the contributions of $m_h$ and $m_H$
to $\Delta\rho$. However, for $m_{H^{\pm}}$ is around $m_H$, the
contributions to $\Delta\rho$ from $(m_h,~m_{H^{\pm}})$ and
$(m_A,~m_{H^{\pm}})$ loops can be canceled by the $(m_h,~m_{H})$ and
$(m_A,~m_{H})$ loops. Thus $m_A$ is allowed to vary from 70 GeV to
700 GeV for $m_{H^{\pm}}$ around 100 GeV.

The contributions of the light CP-even Higgs boson to $\chi^2$ can
be sizably suppressed for $m_h<$ 125 GeV. Therefore, we classify the
surviving samples into groups: 20 GeV $\leq m_H<$ 125 GeV and 125
GeV $\leq m_H\leq$ 125.5 GeV. In Fig. \ref{b-mixing}, the two groups
of surviving samples are projected on the planes of mixing angles.
Fig. \ref{b-mixing} (a) shows that the samples with $\tan\beta>$ 20
require $\sin(\beta-\alpha)$ to approach to 0. Fig. \ref{b-mixing}
(b) shows that, for $m_h<$ 125 GeV, $\theta_d$ can loose the
constraints on $\sin(\beta-\alpha)$ sizably. For example, for
$\theta_d\simeq0$ (Type-II and Flipped 2HDMs), $\sin(\beta-\alpha)$
is allowed to vary from -0.1 to 0.06. While for
$\theta_d\simeq\frac{\pi}{2}$ (Type-I and Lepton-specific 2HDMs),
$\sin(\beta-\alpha)$ is allowed to vary in the range of
$-0.5\thicksim0.44$.  Further, Fig. \ref{b-mixing} (c) shows that
$\theta_l\simeq0$ (Type-II and Lepton-specific 2HDMs) also gives the
strong constraints on $\sin(\beta-\alpha)$, -0.18
$\leq\sin(\beta-\alpha)\leq$ 0.12.

Similar to scenario A,  Figs. \ref{b-mixing} (d) and (e) show that,
although the surviving samples favor 1 $<\tan\beta<$ 7, the value of
$\chi^2$ can be smaller than SM for a large $\tan\beta$ when
$\theta_d$ and $\theta_l$ have the proper large values. Even for
$\tan\beta=41$, the value of $\chi^2$ can be smaller than SM for
$\theta_d=0.7$ and $\theta_l=2.1$. Fig. \ref{b-mixing} (f) shows
that the samples with smaller than SM are in the range of 0.5
$<\theta_d<$ 2 and 0.5 $<\theta_l<$ 2.2. The minimal value of
$\chi^2$ (81.5) appears at $\theta_d=$ 1.8 and $\theta_l=$ 1.1.

In Fig. \ref{b-coup}, the surviving samples with 20 GeV $\leq m_h<$
125 GeV are projected on the planes of Higgs couplings. Similar to
scenario A, for the $HVV$ coupling with the small absolute value,
the $Hb\bar{b}$ coupling by suppressed properly is required to
obtain enough large $Br(h\to ZZ^*)$ and $Br(h\to \gamma\gamma)$. The
constraints on $h\tau\bar{\tau}$ is much more weaken than
$hu\bar{u}$ and $hd\bar{d}$. For the samples with smaller $\chi^2$
than SM,  there is the same sign for the heavy CP-even Higgs
couplings to fermions and gauge bosons. Compared to SM, the $HVV$,
$Hu\bar{u}$ and $Hd\bar{d}$ couplings are suppressed, and the
suppressions are allowed to be as low as 0.94, 0.86 and 0.77,
respectively. However, the $Hl\bar{l}$ coupling can be allowed to
have a $10\%$ enhancement, or $17\%$ suppression.

%%%%%%%%%%%%%%%%%%%%
\begin{table}
\caption{The detailed information of the four samples with the
minimal values of $\chi^2$ in the scenario A (125.5 GeV $\leq m_H<$
128 GeV and 128 GeV $\leq m_H \leq900$ GeV) and scenario B (20 GeV
$\leq m_h <125$ GeV and 125 GeV $\leq m_h\leq125.5$ GeV). Where
$R_{Au\bar{u}}$, $R_{Ad\bar{d}}$ and $R_{Al\bar{l}}$ are from the
interactions, $\frac{m_f}{v}R_{Af\bar{f}}$ $A\bar{f}\gamma^5f$ with
$f=u,~d,~l$.}
\renewcommand{\arraystretch}{0.96}
  \setlength{\tabcolsep}{2pt}
  \centering
  \begin{tabular}{|c|c|c|c|c|}
    \hline
     &~~  scenario A ~~&~~ scenario A~~ &~~ scenario B~~&~~ scenario B~~\\
     \hline
     $m_h$ (GeV)
     & 125.5 & 125.5 &  99.3  &  125.4
     \\
     $m_H$ (GeV)
     & 126.1 & 259.9 &  125.5  &  125.5
     \\
     $m_A$ (GeV)
     & 258.9 & 217.4 &  598.8  &  342.3
     \\
     $m_{H^{\pm}}$ (GeV)
     & 139.1 & 242.8 &  612.1  &  347.1
     \\
     \hline
     $m_{12}^2$ (GeV)
      & 900  & 10000  &  0.01  &  900
      \\
     $\sin(\beta-\alpha) $
     & 0.172 & -0.973 &  0.222  &  -0.042
     \\
     $\tan\beta$
     & 16.48 & 3.57 & 3.91  &  17.07
     \\
     $\theta_d$
     & 1.71 & 1.63 &  1.78  &  1.53
     \\
     $\theta_l$
     & 1.93 & 1.03 &  1.06  &  1.30
     \\
     \hline
     $~~~\chi^2~~~$
     & 83.3 & 81.0 &  81.5  &  83.0
     \\
     \hline
     $R_{hVV}$
     & 0.172 & -0.973 &  0.222  &  -0.042
     \\
     $R_{hu\bar{u}}$
     & 0.231 & -0.909 &  0.472  &  0.016
     \\
     $R_{hd\bar{d}}$
     & 0.371 & -0.895 &  0.702  &  -0.020
     \\
     $R_{hl\bar{l}}$
     & 0.608 & -1.04 &  -0.033  &  -0.260
     \\
     \hline
     $R_{HVV}$
     & 0.985 & 0.229 &  0.975  &  0.999
     \\
     $R_{Hu\bar{u}}$
     & 0.975 & 0.502 &  0.918  &  1.002
     \\
     $R_{Hd\bar{d}}$
     & 0.950 & 0.561 &  0.866  &  1.000
     \\
     $R_{Hl\bar{l}}$
     & 0.909 & -0.040 &  1.033  &  0.990
     \\
     \hline
     $R_{Au\bar{u}}$
     & -0.061 & -0.280 & -0.256  & -0.059
     \\
     $R_{Ad\bar{d}}$
     & 0.202 & 0.341 & 0.491  &  0.022
     \\
     $R_{Al\bar{l}}$
     & 0.443 & -0.277 & -0.262  & -0.217
     \\
     \hline
  \end{tabular}
\label{minimal}
\end{table}

In Table \ref{minimal} we present the detailed information for the
four samples with the minimal values of $\chi^2$ in the scenario A
(125.5 GeV $\leq m_H<$ 128 GeV and 128 GeV $\leq m_H \leq900$ GeV)
and scenario B (20 GeV $\leq m_h <125$ GeV and 125 GeV $\leq
m_h\leq125.5$ GeV). For the four cases, $\theta_d$ and $\theta_l$ of
the samples with the minimal $\chi^2$ are in the ranges of
$1.5\thicksim1.8$ and $1.0\thicksim2.0$. For the scenario A with 128
GeV $\leq m_H \leq900$ GeV and scenario B with 20 GeV $\leq m_h
<125$ GeV, the absolute values for the 125.5 GeV Higgs couplings to
$VV$ approach to SM, and the couplings to $u\bar{u}$ and $d\bar{d}$
have around $10\%$ suppressions compared to SM. The minimal $\chi^2$
values of the two cases are respectively 81.0 and 81.5, which are
marginally smaller than SM value (82.2). This implies that the A2HDM
can provide marginally better fit to the observed Higgs signals than
SM at the expense of additional parameters. Similarly, the minimal
dilaton model can not provide much better fit to LHC and Tevatron
Higgs data than SM \cite{dilaton}. The fit given by little Higgs
models at most approaches to SM for very large scale $f$
\cite{lh-1,lh-2,lh-3}, while Next-to-Minimal Supersymmetric Standard
Model \cite{susy-1,susy-2,susy-3} can give much better fit than SM.

After Moriond 2013, the CMS diphoton data has changed drastically,
which is no longer enhanced. In addition to the four typical 2HDMs,
the Higgs data after Moriond 2013 have been used to examine the
A2HDM in Refs. \cite{2h-11,2h-10,a2hw-3,a2hw-6}. Refs. \cite{a2hw-6}
assumes the both Higgs doublet fields ($\Phi_1$ and $\Phi_2$) to
couple to the up-type quarks, down-type quarks and charged leptons
with aligned Yukawa matrices. However, Refs.
\cite{2h-11,2h-10,a2hw-3} and this paper use a freedom to eliminate
the coupling of up-type quarks to $\Phi_1$. In our discussions, we
consider more relevant theoretical and experimental constraints than
Refs. \cite{2h-11,2h-10,a2hw-3}. Our paper shows that the
theoretical constraint from perturbativity disfavors a large
$\tan\beta$ much more visibly than Ref. \cite{2h-10}. In our
analysis, we consider the 73 Higgs signal strengths observables from
ATLAS, CMS, CDF and D0 collaborations as well as the four Higgs mass
measurements from ATLAS and CMS, which are more than Refs.
\cite{2h-11,2h-10,a2hw-3}. The $\textsf{HiggsSignals-1.1.0}$ is
employed to takes into account the signal efficiencies, experimental
mass resolution and uncertainties. Our paper shows that the Higgs
couplings to gauge bosons and fermions are not more strongly
constrained than Refs. \cite{2h-11,2h-10,a2hw-3,a2hw-6}. Refs.
\cite{2h-11,a2hw-3} focus on the constraints of the Higgs signals on
the Higgs couplings to gauge bosons and fermions. In addition to
these Higgs couplings, we also give the allowed parameters spaces in
detail, including $\tan\beta$, $\sin(\beta-\alpha)$, $\theta_d$,
$\theta_l$, the neutral and charged Higgs masses, and show
explicitly that the proper $\theta_d$ can loose the constraints on
$\sin(\beta-\alpha)$, $\tan\beta$ and $m_{H^{\pm}}$ sizably. An
interesting finding is that when $\theta_d$ and $\theta_l$ have the
proper large values, the value of $\chi^2$ can be smaller than SM
for a large $\tan\beta$ (even $\tan\beta=41$), although the
2$\sigma$ Higgs data and the relevant theoretical and experimental
constraints favor a small $\tan\beta$.

\section{Conclusion}
In this note, we studied the implications of the latest Higgs
signals on a two-Higgs-doublet model with the alignment of the
down-type quarks and charged lepton Yukawa coupling matrices. In our
analysis, we consider the theoretical constraints from vacuum
stability, unitarity and perturbativity as well as the experimental
constraints from the electroweak precision data, flavor observables
and the non-observation of additional Higgs at collider. We obtained
the following observations:

(i) In the scenario A ($m_h$ is fixed as 125.5 GeV),
$\sin(\beta-\alpha)$ is allowed to be in the range of $-1\thicksim1$
for 125.5 GeV $\leq m_H<$ 128 GeV. For $m_H\geq$ 128 GeV,
$\sin(\beta-\alpha)$ is allowed to be in the ranges of
$0.83\thicksim1$ and $-1\thicksim-0.89$ for the proper $\theta_d$,
but be very close to 1 or -1 for $\theta_d=0$. Also, the mixing
angle $\theta_d$ can loose the constraints on $\tan\beta$ and
$m_{H^{\pm}}$ sizably. Although the surviving samples favor 1
$<\tan\beta<$ 5, the value of $\chi^2$ can be smaller than SM for a
large $\tan\beta$ when $\theta_d$ and $\theta_l$ have the proper
large values. $m_{H^{\pm}}$ is allowed to be below 100 GeV for the
absolute value of $\tan(\beta-\theta_d)$ is very small, and the
samples with the smaller $\chi^2$ than SM favor 0.5
$<\tan(\beta-\theta_d)<$ 0 for $m_{H^\pm}>$ 150 GeV.

(ii) In the scenario B ($m_H$ is fixed as 125.5 GeV),
$\sin(\beta-\alpha)$ is allowed to be in the range of $-1\thicksim1$
for 125 GeV $\leq m_h\leq$ 125.5 GeV, and the minimal absolute value
of $\sin(\beta-\alpha)$ decreases with $m_h$ in principle. The light
CP-even Higgs can be allowed to be as low as 20 GeV for -0.25
$<\sin(\beta-\alpha)\leq$ 0. The constraints of the observed Higgs
signals on the opening decay $H\to hh$ require $\tan\beta$ to be
larger than 4 for $m_h<$ 60 GeV. Similar to scenario A, the mixing
angle $\theta_d$ can loose the constraints on $\sin(\beta-\alpha)$,
$\tan\beta$ and $m_{H^{\pm}}$ sizably. For $m_h<$ 125 GeV,
$\theta_d$ around $\frac{\pi}{2}$ can allow $\sin(\beta-\alpha)$ to
be in the range of $-0.5\thicksim0.44$. Although the surviving
samples favor 1 $<\tan\beta<$ 7, the value of $\chi^2$ can be
smaller than SM for $\tan\beta>$ 40 when $\theta_d$ and $\theta_l$
have the proper large values. $m_{H^{\pm}}$ is allowed to be below
100 GeV for the absolute value of $\tan(\beta-\theta_d)$ is smaller
than 2.5, and the samples with the smaller $\chi^2$ than SM favor
-0.5 $<\tan(\beta-\theta_d)<$ 0 for the large $m_{H^\pm}$.

(iii) The model can provide the marginally better fit to available
Higgs signals data than SM. For $m_h=$ 125.5 GeV, the absolute
values of $hVV$, $hu\bar{u}$ and $hd\bar{d}$ couplings are
respectively allowed to be as low as 0.94, 0.90 and 0.83, and
$\theta_d$ and $\theta_l$ are favored in the ranges of 1 $\thicksim$
2 and 0.5 $\thicksim$ 2.2. For $m_H=$ 125.5 GeV, the $HVV$,
$Hu\bar{u}$ and $Hd\bar{d}$ couplings are respectively allowed to be
as low as 0.94, 0.86 and 0.77,  and $\theta_d$ and $\theta_l$ are
favored in the ranges of 0.5 $\thicksim$ 2 and 0.5 $\thicksim$ 2.2.

\section*{Acknowledgment}
We would like to thank Nazila Mahmoudi, Johan Rathsman and Oscar
St{\aa}l for helpful discussions on the codes SuperIso, 2HDMC and
HiggsBounds. This work was supported by the National Natural Science
Foundation of China (NNSFC) under grant Nos. 11005089, 11105116 and
11175151.


\begin{thebibliography}{99}
\bibitem{cmsh} S. Chatrchyan et al. [CMS Collaboration], \PLB716, 30 (2012).

\bibitem{atlh} G. Aad et al. [ATLAS Collaboration], \PLB716, 1 (2012).

\bibitem{tevh} T.~Aaltonen {\it et al.}  [CDF and D0 Collaborations],
  %``Higgs Boson Studies at the Tevatron,''
  \PRD88, 052014 (2013).

\bibitem{2hdm} T. D. Lee, \PRD8, 1226 (1973).

\bibitem{i-1} H. E. Haber, G. L. Kane and T. Sterling, \NPB161, 493
(1979).

\bibitem{i-2} L. J. Hall and M. B. Wise, \NPB187, 397 (1981).

\bibitem{ii-2} J. F. Donoghue and L. F. Li, \PRD19, 945 (1979).

\bibitem{xy-1}V. D. Barger, J. L. Hewett and R. J. N. Phillips,
\PRD41, 3421 (1990).
\bibitem{xy-2} Y. Grossman, \NPB426, 3 (1994).
\bibitem{xy-3} A. G. Akeroyd and W. J. Stirling, \NPB447, 3 (1995).
\bibitem{xy-4} A. G. Akeroyd, \PLB377, 95 (1996).

\bibitem{xy-5} A. G. Akeroyd, \JPG24, 1983 (1998).
\bibitem{xy-6}
M. Aoki, S. Kanemura, K. Tsumura and K. Yagyu, \PRD80, 015017
(2009).

\bibitem{2h-1} C. -Y. Chen and S. Dawson, \PRD87, 055016 (2013).
\bibitem{2h-2} B. Grinstein and P. Uttayarat, \JHEP1306, 094 (2013) [Erratum-ibid. 1309, 110
(2013)].
\bibitem{2h-3} B. Coleppa, F. Kling, S. Su, arXiv:1305.0002
\bibitem{2h-4} O. Eberhardt, U. Nierste, M. Wiebusch, \JHEP07, 118 (2013).
\bibitem{2h-5} C. -W. Chiang and K. Yagyu, \JHEP1307, 160 (2013).
\bibitem{2h-6}B. Grinstein and P. Uttayarat, \JHEP1306, 094 (2013).
\bibitem{2h-7}C. -Y. Chen, S. Dawson and M. Sher, \PRD88, 015018
(2013).
\bibitem{2h-afb} L. Wang, X.-F. Han, \JHEP1205, 088 (2012).
\bibitem{2h-8}N. Craig, J. Galloway and S. Thomas, arXiv:1305.2424.
\bibitem{2h-9}G. Belanger, B. Dumont, U. Ellwanger, J. F. Gunion and S.
Kraml, \PRD88, 075008 (2013).

\bibitem{2h-11} V. Barger, L. L. Everett, H. E. Logan and G. Shaughnessy, arXiv:1308.0052.
\bibitem{2h-10} D. Lopez-Val, T. Plehn and M. Rauch, \JHEP1310, 134 (2013).

\bibitem{2h-12} S. Choi, S. Jung and P. Ko, \JHEP1310, 225 (2013).
\bibitem{2h-13} S. Chang, S. K. Kang, J. -P. Lee, K. Y. Lee, S. C.
Park and J. Song, arXiv:1310.3374.

\bibitem{a2hm-1} A. Pich, P. Tuzon, \PRD80, 091702 (2009).

\bibitem{a2hw-1} W. Altmannshofer, S. Gori and G. D. Kribs, \PRD86, 115009
(2012).
\bibitem{a2hw-2} Y. Bai, V. Barger, L. L. Everett and G. Shaughnessy, \PRD87, 115013 (2013).
\bibitem{a2hw-3} K. Cheung, J. S. Lee, P.-Y. Tseng, arXiv:1310.3937.
\bibitem{a2hw-4} A. Celis, V. Ilisie, A. Pich,  \JHEP1307, 053 (2013).
\bibitem{a2hw-6} A. Celis, V. Ilisie, A. Pich, arXiv:1310.7941.
\bibitem{a2hw-5} W. Altmannshofer, S. Gori, G. D. Kribs, \PRD86, 115009 (2012).


\bibitem{2h-poten} R. A. Battye, G. D. Brawn, A. Pilaftsis, \JHEP1108, 020 (2011).

\bibitem{htm1} W. Konetschny, W. Kummer, \PLB70, 433 (1977).
\bibitem{htm2} T. P. Cheng, L. F. Li, \PRD22, 2860 (1980).
\bibitem{htm3} L. Wang, X.-F. Han, \PRD87, 015015 (2013).

\bibitem{2hc-1} D. Eriksson, J. Rathsman, O. St{\aa}l, \CPC181, 189-205
(2010); \CPC181, 833-834 (2010).


\bibitem{stupara} J. Beringer {\it et al.} (Particle Data Group), \PRD86, 010001
(2012).

\bibitem{spriso} F. Mahmoudi, \CPC180, 1579-1673 (2009).

\bibitem{bsrdstv} Y. Amhis {\it et al.} [Heavy Flavor Averaging Group],
arXiv:1207.1158

\bibitem{bsuu} R. Aaij {\it et al.} [LHCb Collaboration], \PRL110, 021801
(2013).

\bibitem{butv} http://www.slac.stanford.edu/xorg/hfag/rare/2013/radll/index.html

\bibitem{deltamdms} Particle Data Group, 2013 partial update for the 2014 edition.


\bibitem{deltmq} C. Q. Geng and J. N. Ng, \PRD38, 2857 (1988)
[Erratum-ibid. D 41, 1715 (1990)].

\bibitem{hb-1} P. Bechtle, O. Brein, S. Heinemeyer, G. Weiglein, K. E.
Williams, \CPC181, 138-167 (2010).
\bibitem{hb-2} P. Bechtle, O. Brein, S. Heinemeyer, O. St{\aa}l, T.
Stefaniak, G. Weiglein, K. E. Williams, arXiv:1311.0055.


\bibitem{hs-1} P. Bechtle, S. Heinemeyer, O. St{\aa}l, T. Stefaniak, G. Weiglein, arXiv:1305.1933.
\bibitem{hs-2} P. Bechtle, S. Heinemeyer, O. St{\aa}l, T. Stefaniak, G.
Weiglein, "HiggsSignals-1.1 release note", see
"http://higgsbounds.hepforge.org/HS-$1.1_{-}$ releasenote.pdf."

\bibitem{alt-1} ATLAS Collaboration, ATLAS-CONF-2013-030.
\bibitem{alt-2} G. Aad {\it et al.} [ATLAS Collaboration], \PLB726, 88-119
(2013).
\bibitem{alt-3}  ATLAS Collaboration, ATLAS-CONF-2013-013.
\bibitem{alt-4}  ATLAS Collaboration, ATLAS-CONF-2012-091.
\bibitem{alt-5}  ATLAS Collaboration, ATLAS-CONF-2013-012.
\bibitem{alt-6} ATLAS Collaboration, ATLAS-CONF-2013-034.
\bibitem{alt-7}  ATLAS Collaboration, ATLAS-CONF-2012-160.
\bibitem{alt-8} ATLAS collaboration, ATLAS-CONF-2013-079;
ATLAS-COM-CONF-2013-080.
\bibitem{alt-9}  ATLAS collaboration, ATLAS-CONF-2013-075; ATLAS-COM-CONF-2013-069.

\bibitem{cms-1} CMS Collaboration, CMS-PAS-HIG-13-003.
\bibitem{cms-2}  CMS Collaboration, CMS-PAS-HIG-13-022.
\bibitem{cms-3}  CMS Collaboration, CMS-PAS-HIG-13-017.
\bibitem{cms-4}  CMS Collaboration, CMS-PAS-HIG-13-009.
\bibitem{cms-5}  CMS Collaboration, CMS-PAS-HIG-13-002.
\bibitem{cms-6}  CMS Collaboration, CMS-PAS-HIG-12-015.
\bibitem{cms-7}  CMS Collaboration, CMS-PAS-HIG-13-001.
\bibitem{cms-8}  CMS Collaboration, CMS-PAS-HIG-13-007.
\bibitem{cms-9}  CMS Collaboration, CMS-PAS-HIG-13-004.
\bibitem{cms-10}  CMS Collaboration, CMS-PAS-HIG-13-012.
\bibitem{cms-11}  CMS Collaboration, CMS-PAS-HIG-13-020.
\bibitem{cms-12}  CMS Collaboration, CMS-PAS-HIG-13-019.
\bibitem{cms-13}  CMS Collaboration, CMS-PAS-HIG-13-015.

\bibitem{cdf-1} T. Aaltonen {\it et al.} [CDF Collaboration], \PRD88, 052013 (2013).

\bibitem{d0-1} V. M. Abazov {\it et al.} [D0 Collaboration], \PRD88, 052011 (2013).

\bibitem{cus} E. Cervero and J.-M. Gerard, \PLB712,  255 (2012).


\bibitem{dilaton} J. Cao, Y. He, P. Wu, M. Zhang, J. Zhu,
arXiv:1311.6661.


\bibitem{lh-1} L. Wang, J. M. Yang, J. Zhu, \PRD88, 075018 (2013).
\bibitem{lh-2} X.-F. Han, L. Wang, J. M. Yang, J. Zhu, \PRD87, 055004 (2013).
\bibitem{lh-3} J. Reuter, M. Tonini, \JHEP0213, 077  (2013).

\bibitem{susy-1} J. Cao, Z. Heng, J. M. Yang, J. Zhu, \JHEP1210, 079 (2012).
\bibitem{susy-2} J.-J. Cao, Z.-X. Heng, J. M. Yang, Y.-M. Zhang, J.-Y.
Zhu, \JHEP1203, 086 (2012).

\bibitem{susy-3} J. F. Gunion, Y. Jiang, S. Kraml, \PLB710, 454-459 (2012).


\end{thebibliography}
\end{document}